\documentclass[11pt]{article}
\usepackage{jheppub}

\usepackage{amsmath,amssymb, amsfonts}
\usepackage{graphicx}

\def\braket#1{\mathinner{\langle{#1}\rangle}}

\newcommand{\al}{\alpha'}

\newcommand{\ha}{\frac{1}{2}}

\newcommand{\da}{\dot{\alpha}}

\title{Simple superamplitudes in higher dimensions}
\author[a]{Rutger H. Boels}
\affiliation[a]{II. Institut f\"ur Theoretische Physik Universit\"at Hamburg\\ Luruper Chaussee 149, D- 22761 Hamburg, Germany}

\author[b]{and Donal O'Connell}
\affiliation[b]{Niels Bohr International Academy and DISCOVERY center, Niels Bohr Institute,  Blegdamsvej 17, DK-2100 Copenhagen, Denmark}

\emailAdd{Rutger.Boels@desy.de}
\emailAdd{donal@nbi.dk}


\abstract{We provide simple superspaces based on a formulation of spinor helicity in general even dimensions. As a distinguishing feature these spaces admit a fermionic super-momentum conserving delta function solution to the on-shell supersymmetry Ward identities. Using these solutions, we present beautifully simple formulae for the complete three, four and five point superamplitudes in maximal super Yang-Mills theory in eight dimensions, and for the three and four point superamplitudes in ten dimensional type IIB supergravity. In addition, we discuss the exceptional kinematics of the three point amplitude, and the supersymmetric spinorial BCFW recursion, in general dimensions.}

\begin{document}
\maketitle

\section{Introduction}
Scattering amplitudes are fundamental quantities in quantum field theory. It therefore behooves us to understand these amplitudes as simply as possible. Recent years have seen significant progress in our understanding of scattering amplitudes; this is especially true in the context of four dimensions with massless particles\footnote{See, for example~\cite{Dixon:2011xs} and references therein.}.

A natural question, then, is whether this progress can be extended to higher dimensions. A prime motivation here is the possible application to loop calculation in dimensional regularization for four dimensional theories. Moreover, higher dimensional theories appear naturally in string theory and are therefore intrinsically interesting. Finally, there is the natural curiosity about what distinguishes four from higher dimensions. 

A driving force in recent developments has been the surprising simplicity of scattering amplitudes when expressed in the right variables. The gold standard in this is the result by Parke and Taylor \cite{Parke:1986gb} who expressed the MHV scattering amplitudes in a single term expression using so-called spinor-helicity variables. These variables in four dimensions allow one to express four dimensional on-shell momenta in terms of Weyl spinors. Physically, it allows one to make transformation properties under little group transformations manifest. 

Since supersymmetry is one of the two physical extensions of the Poincare group, it should not be a surprise that spinor-helicity variables have a supersymmetric extension. The Parke-Taylor formula for instance was rewritten by Nair \cite{Nair:1988bq} as a function on an on-shell superspace which unifies all scattering amplitudes related by the on-shell supersymmetry Ward identities in maximally supersymmetric Yang-Mills theory. These Ward identities can be derived \cite{Grisaru:1977px} for four dimensional massless particles directly from the on-shell supersymmetry algebra. 

The spinor helicity method has recently been extended to higher dimensions, first to bosonic variables in six dimensions in  \cite{Cheung:2009dc} then to $D\geq4$ including supersymmetry  in \cite{Boels:2009bv}. A version specific to supersymmetric theories in ten dimensions can be found in  \cite{CaronHuot:2010rj}, while a particular example of a useful superspace in six dimensions was explored in  \cite{Dennen:2009vk}. Several papers studying loop amplitudes using the six dimensional supersymmetric formalism have appeared, e.g. \cite{Bern:2010qa}, \cite{Brandhuber:2010mm} and \cite{Dennen:2010dh}. 

In this paper the formulation of and solutions to the on-shell supersymmetry Ward identities in general higher dimensions are explored. Simple forms of four and three point scattering amplitudes in ten and eight dimensions will be presented in IIB supergravity and maximal super Yang-Mills respectively. These amplitudes have immediate extensions to string theory. Our amplitudes should be compared to the known, lengthy expressions for all the bosonic sub-amplitudes contained in these expressions in the literature. The sheer conciseness of our formulation is remarkable. For example, all the four point amplitudes in type IIB supergravity are described by the following, single term formula:
\begin{equation}
\mathcal{A}_4 = \frac{\delta^{16}(Q)}{s t u} .
\end{equation}
This equation includes, as a component amplitude, the famous four graviton scattering amplitude in ten dimensions involving the $t_8$ tensor. As a further example, all the amplitudes in type IIB string theory which are related by supersymmetry to the four graviton scattering amplitude are given by:
\begin{equation}
\mathcal{A}_4 = \frac{\delta^{16}(Q)}{s t u} \frac{\Gamma(\alpha' s +1)\Gamma(\alpha' t +1)\Gamma(\alpha' u +1) }{\Gamma(1 - \alpha's /2)\Gamma(1 - \alpha't /2) \Gamma(1 - \alpha'u /2)}.
\end{equation}

In addition, we describe on-shell recursion relations which allow one to construct more general, higher point tree level amplitudes. In a companion paper \cite{companion} by one of the present authors the extension of the formalism to include massive super-fields is presented. 

This paper is structured as follows: a unifying notation for spinor helicity in even general dimensions is presented in section \ref{sec:conventions}. In section \ref{sec:susy} this spinor helicity notation is used to construct on-shell superspaces. The general structure of amplitudes as functions on these spaces is derived. This includes the construction of supersymmetric delta function solutions to the Ward identities. Special attention is payed to the exceptional three massless particle case. Several explicit three and four point amplitudes in higher dimensions are found in section \ref{sec:superamps}. On-shell recursion relations for the superamplitudes are constructed in section \ref{sec:BCFW} with a sample application to the five point amplitude. A discussion ensues. Appendix \ref{app:proofsewing} contains a proof of a so-called sewing relation used in loop calculations while appendix \ref{app:threeinsix} discusses the three point amplitude in maximal super Yang-Mills theory in six dimensions.


\section{Spinor helicity in various dimensions}\label{sec:conventions}

This section introduces notation for spinor helicity variables in $D$ dimensions, where $D$ will be taken to be even. The notation aims at maximum compatibility between different dimensions. A very useful reference for information on spinors in various dimensions (and signatures) is~\cite{VanProeyen:1999ni}.

\subsubsection*{Chiral spinor notation}
Capital roman indices indicate Weyl spinors and run from $1$ to $\mathcal{D} \equiv 2^{D/2-1}$. Primed and unprimed indices indicate chiral and anti-chiral Weyl spinors. It is convenient to choose a chiral representation of the gamma matrix algebra, i.e.
\begin{equation}
\Gamma^{\mu} = \left(\begin{array}{cc} 0  & \sigma^{\mu,BA'} \\ \bar{\sigma}^{\mu}_{B'A} & 0 \end{array} \right) \, .
\end{equation}
Every Dirac spinor can be expressed in terms of two Weyl spinors as
\begin{equation}
\psi = \left(\begin{array}{c} \lambda^{A} \\ \tilde \lambda_{A'} \end{array} \right)  \, .
\end{equation}
The gamma matrix algebra fixes
\begin{equation}\label{eq:chiralDiracalg}
\begin{array}{rl}
\sigma^{\mu,AA'} \bar{\sigma}^{\nu}_{A'B}  + \sigma^{\nu,AA'} \bar{\sigma}^{\mu}_{A'B} &= 2 \,\eta^{\mu\nu}\, \delta^{A}{}_{B} \\
\bar{\sigma}^{\mu}_{A'A} \sigma^{\nu,AB'}  + \bar{\sigma}^{\mu}_{A'A} \sigma^{\nu,AB'} & = 2 \, \eta^{\mu\nu}\, \delta_{A'}{}^{B'}   \, ,
\end{array}
\end{equation}
while rotations on the chiral spinors are generated by
\begin{align}
\frac{i}{4} \left[\sigma^{\mu,AA'} \bar{\sigma}^{\nu}_{A'B}  - \sigma^{\nu,AA'} \bar{\sigma}^{\mu}_{A'B}  \right] & \equiv (\sigma^{\mu\nu})^{A}{}_{B} \\
\frac{i}{4} \left[ \bar{\sigma}^{\mu}_{A'A} \sigma^{\nu,AB'}  - \bar{\sigma}^{\mu}_{A'A} \sigma^{\nu,AB'}  \right] & \equiv  (\sigma^{\mu\nu})_{A'}{}^{B'}   \, .
\end{align}
In particular, the Weyl spinors $\lambda^{A}$ and $\tilde \lambda_{A'}$ transform as
\begin{equation}
\delta \lambda^{A} = (\sigma^{\mu\nu})^{A}{}_{B} \lambda^B \qquad \delta \tilde \lambda_{A'} = (\sigma^{\mu\nu})_{A'}{}^{B'} \tilde \lambda_{B'}  \, .
\end{equation}
There is always a charge conjugation matrix such that 
\begin{equation}
C \Gamma^{\mu} C^{-1} =  - \left(\Gamma^{\mu}\right)^T  \, ,
\end{equation}
and in particular
\begin{equation}
C \Sigma_{\mu \nu} C^{-1} = - \Sigma_{\mu \nu}^T  \, ,
\end{equation}
holds. The charge conjugation matrix can be used to construct natural Lorentz invariants for two Dirac spinors $\psi$ and $\lambda$,
\begin{equation}
\lambda^T C \psi  \, .
\end{equation}
In our chiral basis the matrix $C$ can be written as 
\begin{equation}
C = \left(\begin{array}{cc} \Omega_{BA}  & 0 \\  0 & \Omega^{B'A'} \end{array} \right),  \qquad {D\,=\,4k+4}  \, ,
\end{equation}   
in dimensions $4,8,\ldots$ and as 
\begin{equation}
C = \left(\begin{array}{cc} 0 & \Omega_{B}{}^{A'}  \\   \Omega^{B'}{}_{A}  & 0 \end{array} \right),  \qquad {D\,=\,4k+2}  \, ,
\end{equation}
in dimensions $6,10,\ldots$. These matrices have natural inverses
\begin{align}
\Omega_{BA} \Omega^{AC} &= \delta_{B}{}^{C} \quad \Omega_{B'A'} \Omega^{A'C'} = \delta_{B'}{}^{C'} \\  
\Omega_{B}{}^{A'}  \Omega_{A'}{}^{C} &= \delta_{B}{}^{C}  \quad  \Omega^{B'}{}_{A}  \Omega^{A}{}_{C'} =  \delta^{B'}{}_{C'}   \, ,
\end{align}
and can be used either to raise and lower chiral spinor indices as 
\begin{equation}
\left. \begin{array}{rl} \lambda_{A} & \equiv \lambda^{B} \Omega_{BA} \\ \tilde \lambda^{A'} & \equiv \tilde \lambda_{B'} \Omega^{B'A'} \end{array} \right. \qquad {D\,=\,4k+4}  \, ,
\end{equation}
or relate chiral and anti-chiral solutions to the Dirac equation as
\begin{equation}
\left. \begin{array}{rl} \lambda_{A} & \equiv \tilde \lambda_{A'} \Omega^{A'}{}_{A} \\ \tilde \lambda^{A'} & \equiv  \lambda^{A} \Omega_{A}{}^{A'} \end{array} \right. \qquad {D\,=\,4k+2}  \, .
\end{equation}
These two possibilities simply state whether or not the Weyl representation is self-conjugate or complex-conjugate. One can choose charge conjugation matrices which obey
\begin{equation}\label{eq:sympropchargeconj}
\left\{ \begin{array}{lcl}
C^T = - C & \quad &  D= 4,10 \\ 
C^T = C & \quad & D= 6, 8  \, ,
\end{array} \right.
\end{equation}
in the number of dimensions relevant for this article~\cite{VanProeyen:1999ni}\footnote{Note that one can always define a matrix $\tilde{C} = C \Gamma_{*}$ for which $\tilde{C} \Gamma^{\mu} \tilde{C}^{-1} =  \left(\Gamma^{\mu}\right)^T$. In $D=4 k + 2$ this would lead to opposite symmetry properties for $\tilde{C}$.}. The Lorentz invariant spinor contractions can be written in this notation as
\begin{equation}
\lambda_{A} \psi^A \equiv [\lambda \psi] \qquad \lambda^{A'} \psi_{A'} \equiv \langle \lambda \psi \rangle  \, .
\end{equation}
The definition of these spinor products either equals
\begin{equation}
 [\lambda \psi] = \lambda^B \Omega_{BA} \psi^A \quad \langle \lambda \psi \rangle =  \lambda_{B'} \Omega^{B'A'} \psi_{A'} \qquad D = 4k+4  \, ,
\end{equation}
or 
\begin{equation}
[\lambda \psi] = \lambda_{A'} \Omega^{A'}{}_{A} \psi^A \quad \langle \lambda \psi \rangle =  \lambda^{A} \Omega_{A}{}^{A'} \psi_{A'} \qquad {D\,=\,4k+2}  \, ,
\end{equation}
depending on dimension. Independent of dimension the conjugate spinors transform as
\begin{equation}
\delta \lambda_{A} = - (\sigma^{\mu\nu})_{A}{}^{B} \lambda_B \qquad \delta \lambda^{A'} = - (\sigma^{\mu\nu})^{A'}{}_{B'} \lambda^{B'}  \, ,
\end{equation}
which makes the above brackets Lorentz invariants. From the properties of the charge conjugation matrix \eqref{eq:sympropchargeconj} it follows that 
\begin{equation}
\begin{array}{lcc}
D=4 & \qquad & [\lambda \psi] = - [ \psi \lambda] \qquad  \langle \lambda \psi \rangle = - \langle \psi \lambda \rangle \\ 
D=6 & \qquad & [\lambda \psi] =  \langle \psi \lambda \rangle \\
D=8 & \qquad & [\lambda \psi] =  [ \psi \lambda] \qquad  \langle \lambda \psi \rangle =  \langle \psi \lambda \rangle \\ 
D=10 & \qquad & [\lambda \psi] = - \langle \psi \lambda \rangle  \, ,
\end{array}
\end{equation}
hold. A  less precise but more number-of-dimension-invariant statement is that among the set
\begin{equation}
\{ \,  [ \psi \lambda] \, , \,[ \lambda \psi ]\, , \,  \langle \lambda \psi \rangle \, , \, \langle  \psi \lambda \rangle\, \}  \, ,
\end{equation}
only two are independent. 

\subsubsection*{Little group transformations}
Spinors of particular interest here arise as the solution to the massless Dirac equation
\begin{equation}\label{eq:choiceofspinorconv}
p_{\mu} \sigma^{\mu,BA'} \lambda_{A',a'} = 0 \qquad p_{\mu} \bar{\sigma}^{\mu}_{A'A}  \lambda^{A,a} = 0  \, ,
\end{equation}
and its conjugate,
\begin{equation}\label{eq:choiceofspinorconvconj}
 \lambda_{B}^{a'}  p_{\mu} \sigma^{\mu,BA'}= 0 \qquad \lambda^{A'}_{a}  p_{\mu} \bar{\sigma}^{\mu}_{A'A}  = 0  \, .
\end{equation}
Here the lower case Roman letters indicate chiral spinor indices of the massless little group in $D$ dimensions. The notation reflects the chiral and anti-chiral Weyl spinors in $D-2$ dimensions. The little group equivalent of $\Omega$ is denoted $\omega$. The spinors appearing here have consistent little group and spinor indices in all even dimensions:
\begin{equation}\label{eq:spinorrelations4k4}
\begin{array}{rl} \lambda_A^{a'} &= \omega^{a'}{}_b\, \Omega_{AB} \lambda^{B,b} \\
\lambda^{A'}_a & =  \omega_a{}^{b'}\, \Omega^{A'B'} \lambda_{B',b'}  \end{array} \qquad  {D\,=\,4k+4}  \, ,
\end{equation}
or
\begin{equation}\label{eq:spinorrelations4k2}
\begin{array}{rl} \lambda_A^{a'} &= \omega^{a'b'} \, \Omega_{A}{}^{B'} \lambda_{B',b'} \\
\lambda^{A'}_a & =  \omega_{ab} \, \Omega^{A'}{}_{B} \lambda^{B,b}  \end{array} \qquad  {D\,=\,4k+2} \, ,
\end{equation}
holds depending on dimension.

The $\lambda$ spinors have vanishing spinor products amongst themselves,
\begin{equation}
[\lambda^{a'} \lambda^a] = 0  \qquad \langle \lambda_a \lambda_{a'} \rangle = 0  \, .
\end{equation}
This follows from the general argument that the little group is a proper subgroup of the Lorentz group. In particular this implies that the little group representation of a product of spinors has to fit into the Lorentz group representation of this product. Here the Lorentz group representation is trivial. It can be checked that the product of spinor representations indicated does not contain a scalar and hence this has to vanish. 

There is a choice of solutions to the Dirac equations \eqref{eq:choiceofspinorconv} and \eqref{eq:choiceofspinorconvconj} such that the massless momentum $p$ can be written 
\begin{equation}\label{eq:spinornorm}
p_{\mu} \sigma^{\mu,BA'} = \lambda^{B, a}  \lambda^{A'}_{a}  \, ,
\end{equation} 
and
\begin{equation}
p_{\mu} \sigma^{\mu}{}_{B'A} = \lambda_{B', a'}  \lambda_{A}{}^{a'}  \, ,
\end{equation}
with the little group index summed. The normalization in equation \eqref{eq:spinornorm} will be imposed throughout this article. This equation follows by the simple observation that the quantity on the right hand side is a little group scalar. The tensor product of the two spinor representations leads in general to a sum over odd space-time forms. However, higher forms than the one form would automatically transform under the little group. Hence the result has to be proportional to a vector orthogonal to the little group transformations which is $p$. 

One way to define the little group indices consistently for all legs is to introduce a light-like vector $q$ for which $q\cdot p \neq 0$ which\footnote{We will assume $q$ is chosen so that $p_i \cdot q \neq 0$ for each leg $i$.} can be used to define the generators of the little group from the three form Pauli-Lubanski tensor,
\begin{equation}\label{eq:defpaulilubvecs}
W_{\mu\nu} = \frac{q^{\rho}}{q\cdot k} k_{[\mu} \Sigma_{\nu\rho]}  \, .
\end{equation}
The light-like vector $q$ can be used to define a set of solutions to the massless Dirac equation,
\begin{equation}
q_{\mu} \sigma^{\mu,BA'} \xi_{A'}^{a} = 0 \qquad q_{\mu} \bar{\sigma}^{\mu}_{A'A}  \xi^{A}_{a'} = 0  \, ,
\end{equation}
or its conjugate with the normalizations 
\begin{equation}\label{eq:spinornorm2}
q_{\mu} \sigma^{\mu,BA'} = \xi^{B}{}_{a'}  \xi^{A',a'}   \, ,
\end{equation}
and
\begin{equation}
\label{eq:xiSpinorNorm}
q_{\mu} \sigma^{\mu}{}_{B'A} = \xi_{B'}{}^{a}  \xi_{A,a}   \, .
\end{equation}

A basis for the little group indices for all legs up to proportionality constants can now be defined using the auxiliary $\xi$ spinors as
\begin{equation}
\lambda^{A,a} \propto p^{AA'} \xi_{A'}^a \qquad \lambda_{A',a'} \propto p_{A'A} \xi^{A}_{a'}  \, ,
\end{equation} 
or equivalently,
\begin{equation}
q_{A'A} \lambda^{A,a}  \propto \xi_{A'}^a \qquad q^{AA'}  \lambda_{A',a'} \propto  \xi^{A}_{a'}  \, .
\end{equation}

The $\lambda$ and $\xi$ spinors when taken together furnish a complete basis of the space of chiral or anti-chiral spinors,
\begin{equation}
\{ \xi^{A}_{a'} ,  \lambda^{A,a}\} \qquad \{ \xi_{A'}^{a} ,  \lambda_{A',a'}  \}  \, .
\end{equation}
This makes explicit that the space of chiral spinors in $D$ dimensions decomposes into the sum of a chiral and an anti-chiral spinor representation in $D-2$ dimensions. Note these $D-2$ dimensions are defined as those orthogonal to the two-plane spanned by $p$ and $q$. 

The inner product of $\xi$ and $\lambda$ spinors are `light-like diagonal': 
\begin{equation}
[\xi_a \lambda^b] = n_{\lambda}  \, \delta_{a}{}^{b} \qquad  \langle \xi^{a'} \lambda_{b'} \rangle = \bar{n}_{\lambda} \,  \delta^{a'}{}_{b'}   \, ,
\end{equation}
as well as
\begin{equation}
[\lambda^{b' } \xi_{a'}] = m_{\lambda} \,  \delta^{b'}{}_{a'}  \qquad  \langle   \lambda_{a}  \xi^{b}\rangle = \bar{m}_{\lambda}   \, \delta_{a}{}^{b}  \, .
\end{equation}
These formula are again dictated by little group transformation properties as a subgroup of the Lorentz transformations. Note that the normalization constants still depend on $\lambda$. Taken together with the normalization of the spinors such as for instance in equation \eqref{eq:spinornorm}
\begin{equation}
\lambda^{A,a} = \frac{p^{AA'} \xi_{A'}^a}{\bar{m}_{\lambda}} \qquad \lambda_{A',a'} = \frac{p_{A'A} \xi^{A}_{a'}}{m_{\lambda}}  \, ,
\end{equation}
follows. From the Dirac algebra
\begin{align}
p^{AA'} q_{A'B}  + q^{AA'} p_{A'B} &= 2 \,p \cdot q\, \delta^{A}{}_{B}  \, ,
\end{align}
holds and hence
\begin{align}
\bar{m}_{\lambda} \lambda^{A,a} \xi_{B,a}  + \bar{n}_{\lambda} \xi^{A}{}_{a'} \lambda_{B}{}^{a'} &= 2 \,p \cdot q\, \delta^{A}{}_{B}  \, ,
\end{align}
which is the completeness relation for the spinors $\lambda$ and $\xi$. Acting with the left and right hand sides on either $\lambda^{A,a}$ and $\lambda^{a'}_A$ (or with $\xi$ spinors) gives the normalization of the spinor products
\begin{equation}\label{eq:spinornorm3}
m_{\lambda} \bar{n}_{\lambda}= 2 p \cdot q  \qquad n_{\lambda} \bar{m}_{\lambda}= 2 p \cdot q  \, .
\end{equation}

Using these formulae one can express a general chiral or anti-chiral spinors in the spinor basis. A generic spinor $Q_{A'}$ can be expressed as
\begin{equation}\label{eq:spinorexpansion}
Q_{A'} = c_{a}(Q) \, \xi_{A'}^{a} + d^{a'}(Q) \,  \lambda_{A',a'}  \, ,
\end{equation}
where the coefficients are 
\begin{equation}
c_a = \frac{\langle \lambda_{a} Q \rangle}{\bar{m}_{\lambda}} \qquad d^{a'} =   \frac{\langle \xi^{a'} Q \rangle }{\bar{n}_{\lambda} }   .
\end{equation}
Similarly for a general spinor $Q^{A}$ one can write 
\begin{equation}
Q^{A} = c^{a'}(Q) \, \xi^{A}_{a'} + d_{a}(Q) \,  \lambda^{A,a}   \, ,
\end{equation}
where the coefficients are 
\begin{equation}
c^{a'}(Q) = \frac{[ \lambda^{a'} Q]}{m_{\lambda}} \qquad d_{a}(Q)  =   \frac{[ \xi_{a} Q ] }{n_{\lambda}}  \, .
\end{equation}
These expansions will be useful below. Note that different choices of $q$ will lead to (numerically) different bases for the spinor space: these are related by a little group transformation.

\subsubsection*{Vectors to spinors and vice-versa}
The heart of the spinor helicity method is a complete dictionary between vectors, tensors, etc., and spinors. To illustrate a general reasoning used throughout this article, consider
\begin{equation}
\label{eq:defOfT}
(T^{\mu})_a{}^{b} = \lambda^{A'}_a \sigma^{\mu}_{A' A} \lambda^{A,b}  \, .
\end{equation}

This object transform under (space-time) Lorentz rotations as a vector. Clearly, $T$ is a tensor product of two spinors of the little group. We can Clebsch out the various $n$ forms of the little group using $\gamma$ matrices as usual. Thus, $(T^{\mu})_a{}^{b}$ transforms as a space-time vector and a little group $0,2,4,\ldots$ form. But note that the little group is a subgroup of the Lorentz group. It follows that the little group representation must be contained in the Lorentz group representation. Therefore, the representations are not independent. Since $T$ is a Lorentz vector, it must be a little group scalar as the other little group representations in the tensor product are too large. Moreover, since it is a little group scalar, the vector must be proportional to $p$. That is,
\begin{equation}\label{eq:clebschmepsch}
(T^{\mu})_a{}^{b} =  2 \delta_a^b \, p^{\mu}  \, .
\end{equation}
The normalization follows from the spinors normalized as in equations~\eqref{eq:spinornorm},~\eqref{eq:xiSpinorNorm} and~\eqref{eq:spinornorm3}.

There is a nice interaction between Clebsching in space-time and in the little group which is worth exploring. For example, consider 
\begin{equation}
\sim \lambda^{a'} \sigma^{\mu \nu} \lambda^a  \, .
\end{equation}
This object is a space-time two-form. From its little group structure, this must be a sum of little group $1,3,\ldots$ forms. By considering dimensions and the fact that the little group representation has to be a part of the space-time two form representation, only the little group vector part remains. It therefore has the transformation properties of the (free) field strength tensor of a vector boson. That is,
\begin{equation}
\lambda^{a'} \sigma^{\mu \nu} \lambda^a \propto (p^\mu \epsilon^{\nu,n} - p^\nu \epsilon^{\mu,n}) \gamma_n^{a' b}  \, ,
\end{equation}
where $\gamma$ is a chiral part of the little group gamma matrix algebra. Choosing a gauge $v$ so that $v \cdot \epsilon^n = 0$ and $p \cdot v \neq 0$, it is easy to see that
\begin{equation}
\epsilon^{\mu,n} \gamma_n^{a' a} \propto \frac{\lambda^{a'} \sigma^\mu v \!\!\! / \lambda^{a} }{2 p \cdot v}  \, .
\end{equation}
A similar result follows from considering the conjugate expression
\begin{equation}
\sim \lambda_{b'} \bar{\sigma}^{\mu \nu} \lambda_{a'}  \, ,
\end{equation}
which leads to 
\begin{equation}
\bar{\epsilon}^{\mu,n} \gamma_{n a a'} \propto \frac{\lambda_{a} \bar{\sigma}^\mu v \!\!\! / \lambda_{a'} }{2 p \cdot v}  \, .
\end{equation}

To compute the normalization of these objects, it is convenient to consider polarization objects $\epsilon^{\mu a' b}$ and $\epsilon^\mu_{a b'}$, defined as
\begin{align}\label{eq:polvecc}
\epsilon^\mu_{a b'} &= \frac{\lambda_{a} \bar{\sigma}^\mu v \!\!\! / \lambda_{b'} }{2 p \cdot v} \\
\epsilon^{\mu a' b} &=  \frac{\lambda^{a'} \sigma^\mu v \!\!\! / \lambda^{b} }{2 p \cdot v}.
\end{align}
This objects transform as spacetime vectors. Consideration of the little group representations available shows that they also transform as a vector of the little group. Thus,
\begin{equation}
\epsilon^\mu_{a a'} = \chi \gamma^m_{a a'} \epsilon^\mu_m, \qquad
\epsilon^{\mu a' a} = \chi \gamma^{a' a m} \bar \epsilon_m^\mu ,
\end{equation}
where $\chi$ is a normalization factor. It is straightforward to demonstrate that
\begin{equation}
\epsilon^\mu_{a b'} \epsilon^{\nu b' b} + \epsilon^\nu_{a b'} \epsilon^{\mu b' b} = 2 \left[ \eta^{\mu \nu} - \frac{p^\mu v^\nu + v^\mu p^\nu}{p \cdot v} \right] \delta_a^b.
\end{equation}
It follows that the choice $\chi = 1/d$ ensures that the polarization vectors satisfy
\begin{equation}
\epsilon^{\mu,n} \bar{\epsilon}^{\nu}_n = \eta^{\mu\nu} - \frac{p^{\mu} q^{\nu} + q^{\mu} p^{\nu} }{q \cdot p}  \, ,
\end{equation} 
Thus, the correctly normalized vectors are 
\begin{equation}
\epsilon^{\mu,n} = \gamma^{n}_{b a'} \frac{\lambda^{a'} \sigma^\mu v \!\!\! / \lambda^{b} }{2 d p \cdot v}  \, .
\end{equation}

\subsubsection*{Further useful formulae}
A further set of formulas useful in this article involve the totally anti-symmetric tensor
\begin{equation}
\epsilon_{A_1 \ldots A_{\mathcal{D}}}  \, .
\end{equation}
Several useful results involving this follow from the relation 
\begin{equation}
 \epsilon_{A_1 \ldots A_{\mathcal{D}}}   \lambda^{A_1,1} \ldots \lambda^{A_{\mathcal{D}/2},d} = \epsilon_{a'_1 \ldots a'_{d}}  \lambda_{A_{\mathcal{D}/2+1}}^{a'_1} \ldots \lambda_{A_{\mathcal{D}}}^{ a'_{d} }.
\end{equation}
This follows from the observation that the spinors $\{\lambda_{A}^{a'}, \xi_{Aa}\}$ are a basis of the spinor space, and so we may build a basis for the tensor product space from products of these spinors. Expressing the quantity $\epsilon_{A_1 \ldots A_{\mathcal{D}}}   \lambda^{A_1,1} \ldots \lambda^{A_{\mathcal{D}/2},d}$ in terms of such a basis, it is easy to see that the only term contributing is as shown. It follows that
\begin{equation}\label{eq:loweringspinorsepsilon}
 \epsilon_{a_1 \ldots a_{d}} \epsilon_{A_1 \ldots A_{\mathcal{D}}}   \lambda^{A_1,a_1} \ldots \lambda^{A_{d,a_{d}}} =  d ! \,\, \epsilon_{a'_1 \ldots a'_{d}}  \lambda_{A_{d+1}}^{a'_1} \ldots \lambda_{A_{\mathcal{D}}}^{ a'_{d} } \, .
\end{equation}
where $\mathcal{D} = 2^{D/2 -1}$ is the chiral spinor dimension in $D$ dimensions and $d$ is the same in $D-2$ dimensions, $d = \mathcal{D}/2$. A special case of this relation, which will be of use to us later in the discussion, is
\begin{equation}\label{eq:spinorsleadingtoinnprod}
\epsilon_{A_1 \ldots A_{\mathcal{D}}}  1^{A_1,a_1} \ldots 1^{A_{d,a_{d}}}  2^{A_{d+1,b_1}} \ldots 2^{A_{\mathcal{D},  b_{d} }}  
 =  \epsilon^{a_1 \ldots a_{d}} \epsilon^{b_1 \ldots b_{d}} (2 p_1 \cdot p_2)^{\mathcal{D}/4} \, ,
\end{equation}
where $1^{A,a}$ and $2^{A,a}$ denote the spinors associated to two massless momenta $p_1$ and $p_2$.

\subsection{Comments on odd dimensions and existing spinor helicity formulations}

\subsubsection*{Odd dimensions}
Spinor helicity in odd dimensions can be obtained in various related ways. A first method is to formulate everything in terms of Dirac spinors in one dimension lower and adjoin $\Gamma_*$ as the extra-dimensional gamma matrix. Another possibility is to restrict the momentum of a particle in one dimension higher to vanish in a particular direction. In the first case the chiral and anti-chiral spinors of the massless little group in $D-1$ (even) dimensions combine into the massive little group in $D$ dimensions. In the other case the massless little group in, $D+1$ dimensions contains the massless little group in $D$ dimensions. In a purely $D$ dimensional language, one works with Dirac matrices rather than $\sigma$ matrices. Indices of these matrices can straightforwardly be raised and lowered using the charge conjugation matrix. The $D-2$ dimensional little group is similarly straightforward.

\subsubsection*{Relation to spinor helicity constructions in the literature}
In four dimensions, the massless little group is $SO(2) \sim U(1)$, which is Abelian. The chirality of a spinor therefore translates immediately into its helicity. To arrive at more standard spinor helicity notation, one should identify capital Roman primed and unprimed indices with Greek dotted and un-dotted indices. The charge conjugation matrix in the chiral representation of the gamma matrix algebra reads for Dirac spinors
\begin{equation}
\Omega = \left(\begin{array}{cc} \epsilon_{\alpha \beta} & 0 \\ 0 & \epsilon^{\dot{\alpha} \dot{\beta}}  \end{array} \right) \, .
\end{equation}
so that the chiral spinor indices are raised and lowered by the two dimensional anti-symmetric tensor $\epsilon$. The momenta obey
\begin{equation}
p^{\alpha \da} = \lambda^{\alpha} \lambda^{\da} \, ,
\end{equation}
and the $\xi$ spinors constructed from a massless momentum $q$ are  trivial here as the relation
\begin{equation}\label{eq:defhelopd4}
\lambda^{\alpha} = \frac{p^{\alpha \da} \xi_{\da}}{\braket{\lambda \xi}} \, .
\end{equation}
is an identity. The massless momentum $q$ can be used to define the helicity generator on a momentum eigenstate,
\begin{equation}
h \sim \frac{q_{\mu} W^{\mu} }{q \cdot k} \, ,
\end{equation}
where $W^{\mu}$ is the Pauli-Lubanski generator. As is well-known, this generator is independent of the choice of $q$ if it acts on a four dimensional massless state since in this case,
\begin{equation}
W^{\mu} | k \rangle = k^{\mu} | k \rangle \, .
\end{equation}
In more abstract terms, this reflects the fact that the little group is Abelian and one-dimensional: any generator is proportional to any other generator. Note that in terms of the helicity operator defined in equation \eqref{eq:defhelopd4} the spinor $\xi_{\alpha}$ has opposite helicity from the spinor $\lambda_{\alpha}$. The construction of the polarization vectors and spinors in the main text in general dimensions reduces to the well-known spinor helicity method in four, as can be checked explicitly. 

In six dimensions the little group in $SO(4) \sim SU(2) \times SU(2)$. There are two relevant chiral charge conjugation matrices in this dimension: the six dimensional one
\begin{equation}\label{eq:sixDomega}
\Omega^{A}{}_{B'} \quad \Omega^{B'}{}_{A} \, , 
\end{equation}
as well as the four dimensional one for the little group indices, where each SU(2) factor will be given dotted or undotted Greek indices,
\begin{equation}\label{eq:sixDlittleomega}
\omega = \left(\begin{array}{cc} \epsilon_{\alpha \beta} & 0 \\ 0 & \epsilon^{\dot{\alpha} \dot{\beta}}  \end{array} \right) \, .
\end{equation}
The charge conjugation matrices in equation \eqref{eq:sixDomega} can be used to replace all primed by unprimed indices. The chiral spinor indices can be raised and lowered in six dimensions through the epsilon symbol,
\begin{equation}
A^{AB} \epsilon_{ABCD} \propto A_{CD} \, .
\end{equation} 
Multiplying equation \eqref{eq:spinornorm} and the one below it by one of the matrices in equation \eqref{eq:sixDomega} and using the little group omega in \eqref{eq:sixDlittleomega} then gives exactly the expressions found in \cite{Cheung:2009dc}. 

In higher dimensions there are the spinor helicity method proposed in \cite{Boels:2009bv}  and a version of $D=10$ spinor helicity in \cite{CaronHuot:2010rj}. The first of these is in principle equivalent to the above but phrased entirely in a special basis for the little group indices. 


\section{On-shell supersymmetry in even dimensions}\label{sec:susy}
In this section the spinor helicity notation introduced in the previous section is employed to construct on-shell superspaces in general even dimensions. The chiral version of the supersymmetry algebra can be written as
\begin{equation}
\{ \overline{Q}^B,Q^{A'} \} = p^{BA'}  \, ,
\end{equation} 
with the momentum on-shell and massless this can be written as
\begin{equation}\label{eq:susyonshell}
\{ \overline{Q}^B,Q^{A'} \} =  \lambda^{B, a}  \lambda^{A'}_{a}  \, ,
\end{equation} 
by equation \eqref{eq:spinornorm}. 

\subsection{Constructing on-shell superspaces}
It is easy to check that 
\begin{equation}\label{eq:defsuperspacegens}
\overline{Q}^B = \lambda^{B, a}   \frac{\partial}{\partial \eta^a}    \qquad Q^{A'}= \lambda^{A'}_{a} \eta^a \, ,
\end{equation}
is a representation of the on-shell supersymmetry algebra in equation \eqref{eq:susyonshell} since
\begin{equation}
\{\eta^{a},    \frac{\partial}{\partial \eta^b}    \} = \delta^a{}_{b}  \, ,
\end{equation}
holds for fermionic variables. Hence it is natural to promote each leg of an amplitude to a superfield, that is a function of the momentum of this leg $p$ and the set of fermionic coordinates $\eta^{a}$. A field on the space spanned by these coordinates has a natural expansion, 
\begin{equation}
\phi(p,\eta) = \phi^0(p) + \phi_{a}(p) \eta^{a} + \frac{ \phi_{ab}}{2!}\eta^{a} \eta^b \ldots + \left(\eta\right)^{\mathcal{D}/2} \overline{\phi^0}(p) \, .
\end{equation}
where
\begin{equation}
 \left(\eta\right)^{\mathcal{D}/2} = \frac{\epsilon_{a_1 \ldots a_{\mathcal{D}/2}  \eta^{a_1} \ldots  \eta^{a_{\mathcal{D}/2}}}}{\left(\mathcal{D}/2\right)!}
 \end{equation}
The fields in this expansion transform as a anti-symmetrized tensor product of the chiral little group spinor representation, tensored by the little group representation of the field $\phi^0$. In this article the latter will be taken to be the scalar: this is the so-called fundamental multiplet of the massless on-shell supersymmetry algebra. The total number of components of the minimal multiplet thus constructed is $2^{\mathcal{D}/2}$. Hence simply by counting it is natural to study gauge theory in eight and supergravity in ten dimensions with this multiplet. In the latter case the theory is type IIB as the representation constructed here involves a \emph{complex} chiral spinor supercharge\footnote{Although the spinors can be taken to be real (Majorana) in ten dimensions, $\eta^{\dagger} = \frac{\partial}{\partial \eta}$ so the charges are complex.}. 

\subsubsection*{Eight dimensional field content of the fundamental multiplet}
The top state in the representation is an eight dimensional scalar which can be taken to be the $9-10$ component of the gauge field in ten dimensions. As a consequence the fields in the multiplet have a natural $U(1)_R$ charge. The fermionic variables will be taken to have charge $1$, while the top-state $\phi$ will be taken to have charge $2$. This determines the charges of the other fields in the multiplet.  

The dimensions of the little group representations in the eight dimensional fundamental multiplet are
\begin{equation}
\begin{array}{ccc}
\textbf{bosonic      }\quad &  \qquad  & \textbf{fermionic} \\
\begin{array}{ccc}
0 & \quad &   \underline{1}_1 \\[5pt]
2 & \quad &   \underline{6}_0 \\[5pt]
4 & \quad &   \underline{1}_{-1} \\[5pt]
\end{array}
& \qquad & \begin{array}{ccc}
1 & \quad &   \underline{4}_{\ha}\\[5pt]
3 & \quad &   \underline{4}_{-\ha}\\[5pt]
\end{array} 
\end{array} \qquad \, ,
\end{equation}
in terms of representations of $SO(6)$. In this table the first column indicates fermionic weight level and the underlined numbers are the little group representation dimensions. The subscripts indicate $U(1)_R$ charge. This is the $\mathcal{N}=1$ Yang-Mills multiplet in eight dimensions. 

\subsubsection*{Ten dimensional field content of the fundamental multiplet}
The complete fundamental multiplet can be taken to have charge $4$ under the natural $U(1)_R$ R-symmetry of the type IIB algebra, while the fermionic coordinates then have charge $1$. The resulting charge assignment for all the fields in the multiplet identifies the top and bottom parts of the multiplet as natural combinations of the dilaton and axion fields.

In terms of dimensions of representations of $SO(8)$ the field content at each fermionic weight can be calculated to be
\begin{equation}
\begin{array}{ccc}
\textbf{bosonic      }\quad &  \qquad  & \textbf{fermionic} \\
\begin{array}{ccc}
0 & \quad &   \underline{1}_2\\[5pt]
2 & \quad &   \underline{28}_1 \\[5pt]
4 & \quad &   \underline{35}_0 + \underline{35}'_0\\[5pt]
6 & \quad &   \underline{28}_{-1} \\[5pt]
8 & \quad &   \underline{1}_{-2} 
\end{array}
& \qquad & \begin{array}{ccc}
1 & \quad &   \underline{8}_{1\ha}\\[5pt]
3 & \quad &   \underline{56}_{\ha}\\[5pt]
5 & \quad &   \underline{56}_{-\ha}\\[5pt]
7 & \quad &   \underline{8}_{-1\ha}
\end{array} 
\end{array} \qquad \, ,
\end{equation}
where the prime on the second $ \underline{35}$ indicates that it has different Dynkin labels than the other $ \underline{35}$. The subscripts indicate $U(1)_R$ charge. The graviton states are located in the 'middle' at fermionic weight $4$. This result can be obtained by applying \cite{lie} and numerology: from the dimensions of the $SO(8)$ representations which occur in the tensor products of the little group spinor rep there is a unique combination for which the dimensions at each level sum to the appropriate binomial coefficient. 

\subsection*{Clebsch-Gordan coefficients and the ten dimensional superfield}
It is instructive to distinguish between the two different states at the middle of the multiplet. One way to do this is to construct explicit Clebsch-Gordan coefficients for the four-fold anti-symmetrized tensor product,
\begin{equation}
a_{abcd} = h_{ij} \gamma^{ij}_{abcd}  + h_{ijkl}\gamma^{ijkl}_{abcd} \, ,
\end{equation}
where the first is trace-less symmetric in the SO(8) vector indices $ij$ and the second is anti-symmetric in $ijkl$. A more direct route to identification is the observation that the four-fold anti-symmetrized tensor product is either self-dual or anti-self-dual. It is easy to guess that one of these is the graviton, the other the four form with self-dual field strength. To see this is true, consider the following combination of spinors
\begin{equation}\label{eq:spinorversionWeyltensor}
\lambda^{a'}_{[A} \lambda^{b'}_{B}\lambda^{c'}_{C}\lambda^{d'}_{D]} \, ,
\end{equation}
which is anti-symmetrized in the ten dimensional chiral spinor indices. From \cite{lie} it follows this spinor product is either $(0,2,0,0,0)$ or $(1,0,0,2,0)$ in terms of $SO(10)$ representations. The first of these has the symmetries of the linearized Weyl tensor in ten dimensions, while the second is not the four form or its field strength. By contracting with the right Clebsch-Gordan coefficient and multiplying with two gauge vector as above in the vector field example this allows one to project out the polarization vector of the graviton state. From equation \eqref{eq:spinorversionWeyltensor} and equation \eqref{eq:loweringspinorsepsilon} it follows that the graviton state is self-dual in the little group spinor indices.

\subsubsection*{Dimensional reduction from ten and eight to four}
The field content of the superfields constructed above can also be written as representations of $SO(2) \times SO(4)$ and $SO(2) \times SO(6)$ in the eight and ten dimensional cases respectively. The eight dimensional superfield decomposes as
\begin{equation}\nonumber
\begin{array}{ccc}
\textbf{bosonic      }\quad &  \qquad  & \textbf{fermionic} \\
\begin{array}{ccc}
0 & \quad &  (0) \times  \underline{1} \\[5pt]
2 & \quad &  (1) \times  \underline{1} + (0) \times  \underline{4}\\[5pt]
4 & \quad &  (0) \times  \underline{1} \\[5pt]
\end{array}
& \qquad & \begin{array}{ccc}
1 & \quad &   (\ha) \times \underline{2}\\[5pt]
3 & \quad &   (\ha) \times \underline{2}' \\[5pt]
\end{array} 
\end{array}\qquad \, ,
\end{equation}
where the first number is the canonically normalized four dimensional helicity while the second is the dimension of the representation of $SO(4)$. Similarly the four dimensional field content of ten dimensional superfield decomposes as
\begin{equation}\nonumber
\begin{array}{ccc}
\textbf{bosonic}\quad &  \qquad  & \textbf{fermionic} \\
\begin{array}{ccc}
0 & \quad &   (0) \times  \underline{1} \\[5pt]
2 & \quad &   (1) \times  \underline{6} + (0) \times \underline{15} + (0)\times \underline{1}  \\[5pt]
4 & \quad &   (2) \times  \underline{1} + (1) \times( \underline{10} + \underline{6} )+ (0)\times (\underline{20} + \underline{15} + \underline{1})\\[5pt]
6 & \quad &   (1) \times  \underline{6} + (0) \times (\underline{15} +  \underline{1}) \\[5pt]
8 & \quad &   (0) \times  \underline{1} 
\end{array}
& \qquad & \begin{array}{ccc}
1 & \quad &  (\ha) \times \underline{4} \\[5pt]
3 & \quad &  (1 \ha) \times \underline{4}' + (\ha) \times \underline{20}' \\[5pt]
5 & \quad &  (1 \ha) \times \underline{4}' + (\ha) \times \underline{20}' \\[5pt]
7 & \quad &  (\ha) \times \underline{4}
\end{array} 
\end{array}\qquad \, ,
\end{equation}
as can be checked by calculating the weight labels (see \cite{companion}).

\subsection{Supersymmetry Ward identities and fermionic delta functions}
Supersymmetric amplitudes can be written as functions on the on-shell superspace constructed above. More exactly, it is a function on $n$ copies of this space, where $n$ is the number of particles in the amplitude,
\begin{equation}
A_n(\{k_i, \eta_i^a\}) \, .
\end{equation}
The on-shell supersymmetric Ward identities can now be written as the simple constraints
\begin{equation}
Q^{A'} A_n  \,=\, 0\, =\, \overline{Q}^A A_n \, ,
\end{equation}
where the generators are the sum over the generators for each leg separately as given in equation \eqref{eq:defsuperspacegens},
\begin{equation}
Q^{A'} = \sum_i Q^{A'}_i \qquad \overline{Q}^A = \sum_i \overline{Q}^A_i  \, .
\end{equation}
The superspace constructed above is special as it admits a so-called fermionic delta function solution to these constraints. The reason is that the generators $Q^{A'}$ in the representation of equation \eqref{eq:defsuperspacegens} are purely multiplicative. The only non-trivial function annihilated by all generators is for general kinematics,
\begin{equation}
Q A  = 0 \,\, \rightarrow \,\, A \sim Q_1 \ldots Q_{\mathcal{D}} \, .
\end{equation}
This can be phrased more Lorentz invariantly in terms of the so-called fermionic delta function
\begin{equation}
\delta^{\mathcal{D}}(Q) \equiv \frac{1}{\mathcal{D!}} \epsilon_{A'_1 \ldots A'_{\mathcal{D}}} Q^{A'_1} \ldots Q^{A'_{\mathcal{D}}} \, .
\end{equation}
Since this function is a polynomial of Q, the action of  $\overline{Q}$ simply results in an expression proportional to the overall momentum . Hence the general super amplitude can be written as 
\begin{equation}
A = \delta(K) \delta(Q) \tilde{A} \quad \textrm{with} \quad \overline{Q}^{A} \tilde{A} = 0  \, .
\end{equation} 
By conservation of $U(1)_R$ symmetry the function $\tilde{A}$ has fermionic weight 
\begin{equation}\label{eq:fermionicweightcounting}
\sum_i q_i - \mathcal{D} \, ,
\end{equation}
where $q_i$ are the charges of the superfields in the amplitude. In the cases under study this will be either $2$ or $4$ in the eight or ten dimensional cases respectively. This shows that in these theories the fermionic weight is exhausted by the fermionic delta function for four particles. However, for three particles the above delta function would have too much fermionic weight. This phenomenon is well-known from three point amplitudes in four dimensions. The resolution is that special kinematics in the three particle case allows a special solution to the Ward identities. This is the subject of the next subsection. 

\subsection{Exceptional three particle kinematics}
The supermomentum can by equation \eqref{eq:spinorexpansion} always be expanded into a basis of spinors spanned by, say, $1^{A'}{}_a$ and $\xi^{A',a'}$
\begin{equation}\label{eq:expsupermom}
Q^{A'} = c_{a'}(Q) \, \xi^{A',a'} + d^{a}(Q) \, 1^{A'}{}_a  \, ,
\end{equation}
where the coefficients are 
\begin{equation}\label{eq:defofccoef}
c_{a'} = \frac{\langle  Q 1_{a'}\rangle}{\bar{n}_{1} } \qquad d^{a} =   \frac{\langle Q \xi^{a'} \rangle }{\bar{m}_{1}} \, .
\end{equation}

In the special case of three point kinematics there exists little group valued spinors $u(i)^k$ such that
\begin{equation}\label{eq:threepartkintiny}
\langle i_{a} j_{b'}\rangle = \sum_{k=1}^{\mathcal{D}/4} u(i)^{k'}_{a} u(j)_{k',a'}  \, .
\end{equation}
where the sum is over the what will be termed the ``tiny group'' ($SO(D-4)$). Conventions have been chosen for the tiny group spinor indices to reflect those of \eqref{eq:choiceofspinorconv}. Equation \eqref{eq:threepartkintiny} follows simply by considering the plane spanned by the three on-shell momenta. There is a special choice of little group representation which reduces the analysis of the three point function to the well-understood four dimensional case. The tiny group in this case is the group of transformations orthogonal to the four dimensions which contain the momenta. The equation just written is a little group transformation of this. The coefficients $c_{a'}$ in equation \eqref{eq:defofccoef} simplify in the three point case as
\begin{equation}
c_{a'} = \frac{\sum_{k=1}^{\mathcal{D}/4} \left[\eta_2^{a}\, u(2)^{k'}_{a} +\eta_3^{a} \,u(3)^{k'}_{a}\right] u(1)_{k',a'}}{\bar{n}_{1} } \, .
\end{equation}
Hence there are only $\mathcal{D}/4$ independent coefficients $c_{a'}$ in equation \eqref{eq:expsupermom}, while the number of $d^a$ coefficients remains unchanged. From equation \eqref{eq:expsupermom} it then follows that there are only $\frac{3}{4} \mathcal{D}$ independent supercharges $Q^{A'}$. A completely analogous computation yields that there are only $\frac{3}{4} \mathcal{D}$ independent conjugate supercharges $\overline{Q}^{A'}$ as well. 

A Poincare invariant supersymmetric delta function can now be formulated for three particle kinematics as
\begin{equation}\label{eq:threepointdeltafunction}
\delta^{\frac{3}{4}\mathcal{D}}(Q) \equiv \left( \frac{1}{\left(\frac{3}{4}\mathcal{D}\right)!} \epsilon_{A'_1 \ldots A'_{\mathcal{D}}}\right) \, \left(  \xi^{A'_1,a'_1} \ldots \xi^{A'_{\frac{1}{4} \mathcal{D}},a'_{\frac{1}{4} \mathcal{D}}} \right) \, \left( Q^{A'_{\frac{1}{4} \mathcal{D}+1}} \ldots Q^{A'_{\mathcal{D}}} \right) \, L_{a'_1 \ldots a'_{\frac{1}{4} \mathcal{D}}} \, ,
\end{equation}
which involves a function $L$ which transforms in the anti-symmetrized tensor product of the little group anti-chiral spinor representation. This is manifestly annihilated by the operator in equation \eqref{eq:expsupermom}. It is also annihilated by the operator $\bar{Q}$ up to momentum conservation since it is a polynomial function in the variable $Q$. Hence it is invariant under the full super-Poincare algebra. 

From the above expression for the three particle supersymmetric delta function one can derive others. For instance, one can eliminate the $\xi$ spinors from the above expression and work with a function which has one or more free Lorentz indices. Alternatively, one can insert the expansion of equation \eqref{eq:expsupermom} into equation \eqref{eq:threepointdeltafunction} to yield
\begin{equation}
\delta^{\frac{3}{4}\mathcal{D}}(Q) \propto \left[\epsilon_{a_1 \ldots a_{\mathcal{D}/2}} \left(d^{a_1} \ldots d^{a_1} \right) \right] \left[ L^{a'_1 \ldots a'_{\frac{1}{4} \mathcal{D}}}   \tilde{c}^{k'} u(1)_{k',a'_{1}} \ldots \tilde{c}^{k'} u(1)_{k',a'_{\mathcal{D}/4}} \right] \left(\bar{m}_1^{\mathcal{D}/2} \right) \, ,
\end{equation}
where 
\begin{equation}
\tilde{c}^{k'} = \frac{\left[\eta_2^{a}\, u(2)^{k'}_{a} +\eta_3^{a} \,u(3)^{k'}_{a}\right]}{\bar{n}_1} \, ,
\end{equation}
and the indices on the $L$ tensor have been raised using the completely anti-symmetric tensor. The proportionality factor is $\frac{1}{\left(\frac{3}{4}\mathcal{D}\right)!}\left( d/2 \atop d/4 \right)$. 

\subsubsection*{Tiny group freedom}
The tensor $L$ is not uniquely fixed by the constraint that the delta function not vanish. The non-trivial choices for $L$ are related by the tiny group $SO(D-4)$. As shown above the delta function is proportional to
\begin{equation}
\sim \epsilon^{a_1 \ldots a_{\frac{1}{2} \mathcal{D}}} \left(u(1)_{1, a'_1} \ldots u(1)_{\frac{1}{4} \mathcal{D},a_{\frac{1}{4} \mathcal{D} }}\right) \, c_{a'_{\frac{1}{4} \mathcal{D}+1}\ldots a'_{\frac{1}{2} \mathcal{D}}} \, ,
\end{equation}
from which it is seen that $L$ is invariant under
\begin{equation}
L_{a'_1 \ldots a'_{\frac{1}{4} \mathcal{D}}} \rightarrow  L_{a'_1 \ldots a'_{\frac{1}{4} \mathcal{D}}} + u(1)_{j, a'_1}  d_{a'_2 \ldots a'_{\frac{1}{4} \mathcal{D}}} \, ,
 \end{equation}
and its permutations for any little group tensor $d$. Hence effectively the little group indices on $L$ run from $1$ to $\mathcal{D}/4$ and transform as an $\mathcal{D}/4$-fold antisymmetric tensor product of tiny group ($SO(D-4)$) chiral spinor representations. In particular, this shows that $L$ is in a real sense unique which simply reflects well-known four dimensional results. In numerical practice one can simply pick an arbitrary $L$ and then check that the resulting delta function does not vanish.


\section{Explicit superamplitudes in eight and ten dimensions}\label{sec:superamps}
In this section examples of supersymmetric scattering amplitudes in eight and ten dimensions are presented with three and four points. In these cases the supersymmetric delta functions above saturate the fermionic weight of the scattering amplitudes by equation \eqref{eq:fermionicweightcounting}. Hence determining the bosonic proportionality factors by computing component amplitudes determines the amplitudes in these cases. 

As some of our readers may be familiar with six-dimensional spinor helicity the six dimensional three point amplitude in maximal super Yang-Mills theory in the present spinor language has been included in appendix \ref{app:threeinsix}. 

\subsection{Three points}

In the following the color ordered scalar-vector-scalar amplitude in Yang-Mills theory minimally coupled to a scalar will be needed. This amplitude is given in any dimension by
\begin{equation}
A(\bar{\phi}\, g_{aa'} \, \phi) = 2 \, p_{1,\mu} \epsilon^{\mu}_{a a'} \, ,
\end{equation}
In terms of spinor helicity inserting the polarization equation \eqref{eq:polvecc} in $q$-gauge into this expression gives
\begin{equation}
A(\bar{\phi} \, g_{aa'}\, \phi) 	 =   2 \, \frac{  n_1 }{n_{2}} \langle 1_{a}  2_{a'} \rangle \, .
\end{equation}
By momentum conservation
\begin{equation}
1^{A'}_{a} n_1 + 2^{A'}_{a} n_2 + 3^{A'}_{a} n_3 = 0 \, ,
\end{equation}
holds so that
\begin{equation}
\langle 1_{a} 3_{a'} \rangle n_1 + \langle 2_{a} 3_{a'} \rangle  n_2 = 0 \, ,
\end{equation}
and hence
\begin{equation}
\frac{  n_1}{n_2} = \frac{ v^{a a'} \langle 2_{a} 3_{a'} \rangle }{v^{a a'} \langle 1_{a} 3_{a'} \rangle }\, ,
\end{equation}
for any little group tensor $v^{a a'}$ such that $v^{a a'} \langle 1_{a} 3_{a'} \rangle \neq 0$. Moreover,
\begin{equation}
 \frac{ v^{a a'} \langle 2_{a} 3_{a'} \rangle }{v^{a a'} \langle 1_{a} 3_{a'} \rangle } =   \frac{ w^{a a'} \langle 2_{a} 3_{a'} \rangle }{w^{a a'} \langle 1_{a} 3_{a'} \rangle} \, ,
 \end{equation}
holds for two tensors $w$ and $w$. Choose this tensor to read
\begin{equation}
v^{bb'} = w^{b'} \delta_a{}^b \, ,
\end{equation}
so that 
\begin{equation}
A(\bar{\phi}\, g_{aa'}\, \phi) 	 =  2 \frac{ \langle 2_{a} 3_{b'} \rangle w^{b'}}{ \langle 1_{a} 3_{b'} \rangle w^{b'}}  \langle 1_{a}  2_{a'} \rangle \, ,
\end{equation}
which leads by a similar equation as above to 
\begin{equation}\label{eq:correctamp}
A(\bar{\phi} \, g_{aa'}\,  \phi) 	 = 2 \frac{ \langle 2_{a} 3_{b'} \rangle  \langle 1_{b}  2_{a'} \rangle w^{b'}  w^{b}}{ \langle 1_{b} 3_{b'} \rangle w^{b'} w^b}  \, .
\end{equation}
An equivalent form follows in terms of the `$u$' variables of equation \eqref{eq:threepartkintiny} and their natural (pseudo)-inverses $v$ for which,
\begin{equation}
v^a_{l'} u^{k'}_{a} = \delta_{l'}^{k'} \qquad v^{a', l'} u_{k',a'} = \delta^{l'}_{k'} \, ,
\end{equation}
holds. Note that the $v$ spinors are to $u$ as what the $\xi$ spinors are to $\lambda$: in a sense they define the tiny group indices. Now one can choose
\begin{equation}
w^b = v(1)^{b}_{i'} \qquad w^{b'} = v(3)^{b',i'} \, ,
\end{equation}
for any index $i'$ such that the amplitude reads
\begin{equation}
A(\bar{\phi}\, g_{aa'} \,\phi) = \frac{1}{d} u(2)^{i'}_{a} u(2)_{a',i'}  \qquad \textrm{no sum on } i'\, .
\end{equation}
Since this result is independent of the index $i'$ it is natural to sum over it to give
\begin{equation}\label{eq:uversionthreepointamp}
A(\bar{\phi}\, g_{aa'} \,\phi) = 2 \,  u(2)^{i'}_{a} u(2)_{a',i'} \, .
\end{equation}
Note that the calculation of the scalar-vector-scalar amplitude is framed in $D$ (even) dimensions up to this point.

\subsubsection{Yang-Mills theory in eight dimensions}

Specializing the general discussion to eight dimensions, the color-ordered three point amplitude Yang-Mills superamplitude is given by
\begin{equation}\label{eq:threepoint ansatzD8}
\mathcal{A}^{susy} =  X \delta^{6}(Q) \, ,
\end{equation}
where the fermionic delta function is given by \eqref{eq:threepointdeltafunction} and $X$ is a normalization which will be  computed. The three point fermionic delta function contains an undetermined little group anti-symmetric tensor $L_{a'b'}$. Note that $X$ has to be anti-symmetric under interchange of any of the legs on the color ordered amplitude. 

To compute the normalization constant, it suffices to compute a component amplitude which will be taken to be the scalar-vector-scalar amplitude calculated above. Note that in eight dimensions primed and unprimed little group indices may be interchanged,
\begin{equation}
A(\bar{\phi}\, g_{aa'} \,\phi)  = \omega_{a'}{}^{b} A(\bar{\phi}\, g_{ab} \,\phi) 
\end{equation}
The sought-for amplitude can now be isolated from the supersymmetric amplitude in equation \eqref{eq:threepoint ansatzD8} by fermionic integration,
\begin{equation}\label{eq:isolatingYMd8}
A(\bar{\phi} \, g_{ab} \, \phi)  = \int d\eta_1^4 d\eta^a_2 d\eta^b_2 \mathcal{A}^{susy} \, .
\end{equation}
It is easy to perform the integration to yield
\begin{equation}
A(\bar{\phi}\,  g_{ab} \, \phi)  = X \frac{1}{4!} \epsilon_{A'_1 \ldots A_8} \epsilon^{a_1 a_2 a_3 a_4} 1^{A'_1}_{a_1} \ldots 1^{A'_4}_{a_4} 2^{A'_5}_{a} 2^{A'_6}_{b} \xi^{A'_7,a'} \xi^{A'_8,b'} L_{a'b'} \, ,
\end{equation}
which is equal to 
\begin{align}
A(\bar{\phi} \, g_{ab}\,  \phi)  & = X   \epsilon^{a'_1 a'_2 a'_3 a'_4} 1_{A'_1,a'_1} \ldots 1_{A'_4,a'_4}  2^{A'_1}_{a} 2^{A'_2}_{b}  \xi^{A'_3,a'} \xi^{A'_4,b'} L_{a'b'}\\ 
& =  X \bar{n}_1^2  L^{a'b'} 1_{A',a'} 1_{B',b'}   2^{A'}_{a} 2^{B'}_{b} \\ 
& =  - X \bar{n}_1 \bar{n}_3  L^{a'b'} 1_{A',a'} 3_{B',b'}   2^{A'}_{a} 2^{B'}_{b}  \, ,
\end{align}
by equation \eqref{eq:loweringspinorsepsilon} and momentum conservation. Here
\begin{equation}
L^{a'b'} = \epsilon^{a' b' c' d'} L_{c'd'} \, .
\end{equation}
Converting to $u$'s gives
\begin{align}
A(\bar{\phi} \, g_{ab}\,  \phi) & =  - X \bar{n}_1 \bar{n}_3  L^{a'b'} u(1)_{i' a'} u(3)_{j' b'} u(2)^{i'}_{a} u(2)^{j'}_{b} \\
& =   \frac{X}{2} \left( \bar{n}_1 \bar{n}_3  L^{a'b'} u(1)_{i' a'} u(3)^{i'} _{b'} \right) \left( u(2)_{j',a} u(2)^{j'}_{b} \right)
\end{align}
where the last step follows because 
\begin{equation}
L^{a'b'} u(1)_{i' a'} u(3)_{j' b'} = \frac{1}{2} \epsilon_{i'j'} L^{a'b'} u(1)_{k' a'} u(3)_{b'}^{k'}
\end{equation}
since this combination is a little group scalar and the tiny group is a subgroup of this. Hence if $X$ is chosen to be
\begin{equation}
X =  - 4 \left(L^{a'b'} [\xi_{a'} | 1 \!\!\! \slash 3 \!\!\! \slash | \xi_{b'} ]  \right)^{-1} \, ,
\end{equation}
and $L = w^{b'} w^{b}$ then the amplitude of equation \eqref{eq:correctamp} is reproduced, when taking into account that by the action of the little group charge conjugation matrix $\omega$ in this dimension primed and unprimed spinor indices are equivalent. Note that the factor $X$ is anti-symmetric under interchange of any pair of external legs, as required.

\subsubsection{IIB supergravity in ten dimensions}

In ten dimensions, the IIB supergravity three point superamplitude is proportional to the supersymmetric delta function,
\begin{equation}\label{eq:threepoint ansatzD10}
\mathcal{M} = X \mathcal{A}^{susy} =  X \delta^{12}(Q) \, ,
\end{equation}
where $X$ is a normalization factor. This can be determined by computing the amplitude describing two scalars scattering on a graviton. In field theory this result simply follows from taking two copies of the scalar-vector-scalar amplitude in equation \eqref{eq:uversionthreepointamp}. The required amplitude is given through the Kawai-Lewellen-Tye relations \cite{Kawai:1985xq} by
\begin{equation}
A(\bar{\phi} \, h_{a a' b b'}\, \phi) = u(2)^{i'}_{a} u(2)_{a',i'}  u(2)^{j'}_{b} u(2)_{b',j'}  \, .
\end{equation}
The $SO(6)$ tiny group index may be lowered to give
\begin{equation}
A(\bar{\phi}\, h_{a a' b b'}\, \phi) = \epsilon^{i'j'k'l'} u(2)_{i',a} u(2)_{a',j'}  u(2)_{k',b} u(2)_{b',l'}  \, .
\end{equation}
Note in particular that the resulting expression is pair-wise anti-symmetric in the primed and unprimed little group indices. The anti-symmetric product of two chiral or anti-chiral $SO(8)$ indices can be mapped to one another as the resulting Dynkin label of the representation is $(0,1,0,0)$. Hence Clebsch-Gordan coefficients $G_{cd}^{a'b'} $ exist such that 
\begin{align}
A(\bar{\phi}\, h_{a b c d }\, \phi) & = G_{cd}^{a'b'} A(\bar{\phi}\, h_{a a' b b'} \,\phi)  \\
& = \epsilon^{i'j'k'l'} u(2)_{i',a} u(2)_{b,j'}  u(2)_{k',c} u(2)_{d,l'} \, ,
\end{align}
which in particular is completely anti-symmetric in the unprimed indices. The explicit Clebsch-Gordan coefficients are
\begin{equation}
G_{cd}^{a'b'} = \frac{1}{4} (\sigma_{ij})_{cd} (\bar{\sigma}^{ij})^{a'b'} \, ,
\end{equation}
where the sigma's are the rotation matrices. The normalization is fixed by the requirement that this coefficient squares to the identity. The amplitude is, by equation \eqref{eq:loweringspinorsepsilon}, self-dual in the little group indices. Hence it cannot describe the scattering of two scalars of a four form with self-dual field strength. Since the latter amplitude does not exist this is a feature.   

The normalization factor in equation \eqref{eq:threepoint ansatzD10} can be determined by computing from it the just discussed component amplitude. For this the superamplitude is expanded, setting $\eta_3 = 0$ and retaining only terms with eight powers of $\eta_1$ and four powers of $\eta_2$. In principle expanding the four $\eta_{2}$ variables will yield scattering amplitudes for both a graviton as well as a four-form with self-dual field strength. Here the second will drop out, as will be shown shortly. Integrating over the fermionic factors, the amplitude is
\begin{equation}
A(\bar{\phi} \, h_{a b c d }\, \phi) =  X \frac{1}{8!} \epsilon_{A'_1 \cdots A'_{16}} \epsilon^{a_1 \cdots a_8} 1^{A'_1}_{a_1} \cdots  1^{A'_8}_{a_8}  2^{A'_9}_{a} \cdots 2^{A'_12}_{d}  \xi^{A'_{13}, b'_1} \cdots \xi^{A'_{16}, b'_4} L_{b'_1 \cdots b'_4} \, ,
\end{equation}
which is equal to 
\begin{align}
A(\bar{\phi} \, h_{a b c d } \, \phi) &= X  \epsilon^{a'_1 \cdots a'_8} 1_{A'_1, a'_1} \cdots  1_{A'_8, a'_8}  2^{A'_1}_{a} \cdots 2^{A'_4}_{d}  \xi^{A'_{5}}_{b_1} \cdots \xi^{A'_{8}}_{b_4} L^{b_1 \cdots b_4} \\
& =  X  \bar{n}_1^4  L^{a' b' c' d'} 1_{A',a'} 1_{B',b'} 1_{C',c'} 1_{D',d'}   2^{A'}_{a} 2^{B'}_{b} 2^{C'}_{a} 2^{D'}_{b} \, ,
\end{align}
by equation \eqref{eq:loweringspinorsepsilon}. The tensor $L$ is taken to be anti-symmetric. Converting to $u$'s and exploiting anti-symmetry in the $SO(6)$ tiny group gives 
\begin{multline}
A(\bar{\phi} \, h_{a b c d }\,  \phi) = X  \frac{1}{4!} \bar{n}_1^4 \left[ L^{a' b' c' d'} \epsilon^{i'j'k'l'} u(1)_{i',a'} u(1)_{b',j'}  u(1)_{k',c'} u(1)_{d',l'} \right] \\  \epsilon^{i'j'k'l'} u(2)_{i',a} u(2)_{b,j'}  u(2)_{k',c} u(2)_{d,l'} \, .
\end{multline}
As advertised this scattering amplitude only describes the scattering of a graviton, not of the four form with self-dual field strength. Thus, a choice of X in the superamplitude as
\begin{equation}
X = \frac{4!}{ \bar{n}_1^4 \left[ L^{a' b' c' d'} \epsilon^{i'j'k'l'} u(1)_{i',a'} u(1)_{b',j'}  u(1)_{k',c'} u(1)_{d',l'} \right]}  \, ,
\end{equation}
leads to the correct component amplitude. This fixes the normalization of the superamplitude.

\subsection{Four points }
In this subsection scattering amplitudes with four massless legs will be constructed. In this case the scattering amplitudes should have poles. We will show this allows one to basically guess the amplitude, and will confirm this guess by computing a component amplitude.

 \subsubsection{Yang-Mills theory in eight dimensions}
The color ordered four point Yang-Mills amplitude in eight dimensions should have two poles, at $\frac{1}{s}$ and $\frac{1}{t}$. Further, it has to be cyclic which leaves the combination $\frac{1}{s \,t}$. Since the fermionic delta function has mass dimension $4$ the natural guess for the four point scattering amplitude is
\begin{equation}\label{eq:ymindeight}
A_4^{D=8}(1,2,3,4)  = X \frac{\delta^8(K) \delta^8(Q)}{s \, t}  \, ,
\end{equation}
up to dimensionless normalization. In fact, with this guess for the field theory in hand one can immediately extend it to superstring theory (dimensionally reduced to eight dimensions)
\begin{equation}
A_4^{D=8, \textrm{string}}(1,2,3,4)  = X \frac{\delta^8(K) \delta^8(Q)}{s \, t} \left[\frac{\Gamma\left(\al s + 1\right) \Gamma\left(\al t +
1\right) }{\Gamma\left(1- \al u \right)} \right] \, ,
\end{equation}
by multiplying in the Veneziano factor. As a simple check one can calculate the field theory amplitude with four scalars. Since these scalars are the $8-9$ components of the Yang-Mills field in $10$ dimensions one can use the known form of the $4$ point scattering amplitude in $10$ dimensions  (this can be found for instance by taking the field theory limit of the expressions in \cite{Kawai:1985xq}). Since all momenta are chosen to be orthogonal to the $8$ and $9$ direction, the result is very simple
\begin{equation}
A(\phi_1, \phi_2, \overline{\phi}_3, \overline{\phi}_4) = \frac{s}{t} \qquad \textrm{D=8} \, .
\end{equation}

The same amplitude follows from the above superamplitude in equation \eqref{eq:ymindeight} by integrating out all fermions on legs $3$ and $4$. The result is
\begin{align}
A(\phi_1, \phi_2, \overline{\phi}_3, \overline{\phi}_4) &  = X\int d\eta^4_3 d\eta^4_4 A_4^{D=8}(1,2,3,4) \\
										 &  = X\frac{\epsilon_{A'_1 \ldots A'_8} \epsilon^{a_1 a_2 a_3 a_4}\epsilon^{b_1 b_2 b_3 b_4} 3_{a_1}^{A'_1} \ldots 3_{a_4}^{A'_4} 4_{b_1}^{A'_5} \ldots 4_{b_4}^{A'_8}     }{d! d! st}  \\
										 & =  X \frac{s^2 }{st}  = X \frac{s}{t} \, .
\end{align}
The last line follows by equation \eqref{eq:spinorsleadingtoinnprod}. Hence 
\begin{equation}
X=1 \, ,
\end{equation}
which fixes the superamplitude. 

\subsubsection{IIB supergravity in ten dimensions}
The four point supergravity amplitude in any dimension should have three poles, at $\frac{1}{s}$, $\frac{1}{t}$ and $\frac{1}{u}$. Since it has to have total permutation symmetry, this leads to the natural combination $\frac{1}{s\,t\,u}$. Since the fermionic delta function has mass dimension $8$ the natural guess for the four point scattering amplitude is
\begin{equation}\label{eq:gravindten}
A_{4}^{D=10}   =  X \frac{\delta^{10}(K) \delta^{16}(Q)}{s \, t \, u} \, ,
\end{equation}
again up to dimensionless normalization. In fact, with this guess for the field theory in hand one can immediately extend it to superstring theory 
\begin{equation}
A_{4}^{D=10}   =  \frac{\delta^{10}(K) \delta^{16}(Q)}{s \, t \, u} \left[\frac{\Gamma\left(\al s + 1\right) \Gamma\left(\al t +
1\right) \Gamma\left(\al u + 1\right) }{\Gamma\left(1- \frac{(\al s)}{2} \right) \Gamma\left(1- \frac{(\al t )}{2} \right)
\Gamma\left(1- \frac{(\al u)}{2} \right)} \right] \, ,
\end{equation}
by multiplying in the Virasoro-Shapiro factor. 

\subsubsection*{Kawai-Lewellen-Tye relation}
To check the above consider the dimensional reduction of the type IIB supergravity amplitude to eight dimensions where this is a $\mathcal{N}=2$ supersymmetric theory. Since the momenta of the four gravitons span a four dimensional space, this reduction is no restriction on the physical content of the amplitude. 

It is easy to see that the delta function will split by decomposing the $10$ dimensional chiral spinors into eight dimensional chiral and anti-chiral spinors,
\begin{equation}
 \delta^{16}(Q) \rightarrow  \delta^{8}(Q^A)  \delta^{8}(\tilde{Q}_{A'})  \qquad \textrm{dimensional reduction } 10 \rightarrow 8 \, .
\end{equation}
In eight dimensions the Weyl spinors are not related: the two $N=1$ supersymmetries in eight dimensions appear in conjugate representations. This immediately yields an explicitly on-shell supersymmetric version of the Kawai-Lewellen-Tye relations \cite{Kawai:1985xq},
\begin{equation}
A_{4}^{sugra}(\bar{\eta}_a, \eta^a) =  t A_4^{YM}(1,2,3,4)(\bar{\eta}_a) A_4^{YM}(1,3,2,4)(\eta^a) \, ,
\end{equation}
here in eight dimensions (here all amplitudes are stripped of their coupling constants). It is natural to fermionic Fourier transform one of these to get
\begin{equation}
\tilde{A}_{4}^{sugra}(\iota^a, \eta^a) = \int \left[\prod_{i} \left(\prod_{a=1}^4 d\bar{\eta}_{i,a} e^{\bar{\eta}_{i,a} \iota^{a}_i } \right) \right] A_{4}^{sugra} (\bar{\eta}_a, \eta^a) \, .
\end{equation}
which yields 
\begin{equation}
\tilde{A}_{4}^{sugra}(\iota^a, \eta^a) =  t A_4^{YM}(1,2,3,4)(\iota^a) A_4^{YM}(1,3,2,4)(\eta^a) \, ,
\end{equation}
In this superspace the top state of the superfield is the 10-9 dimensional component of the graviton. There is a natural generalization to string theory amplitudes. For higher points it is seen that the supersymmetric delta functions can always be factored out of the KLT relations, just as in four dimensions. Hence after reduction to eight dimensions \eqref{eq:gravindten} reproduces the right component amplitudes. This proves the ten dimensional amplitude.


\section{Supersymmetric on-shell recursion}\label{sec:BCFW}
In this section a supersymmetric version of the Britto-Cachazo-Feng-Witten recursion relations  \cite{Britto:2004ap}, \cite{Britto:2005fq} will be presented which allows calculation of all super amplitudes in principle. An explicit five point example will be presented.

\subsection{Supersymmetric BCFW shifts in D dimensions}
Given two massless momenta $k_1$ and $k_2$ construct a massless momentum $n$ such that
\begin{equation}\label{eq:bcfwconditions}
n \cdot n = n\cdot k_1 = n \cdot k_2 = 0 \, .
\end{equation}
The standard BCFW shift is then
\begin{align}
\label{eq:standardBCFW}
\hat p_1(z) &= p_1  + z \, n \\
\hat p_2(z) &= p_2  - z \,  n \, ,
\end{align}
which conserves bosonic momentum. As in the familiar four dimensional case, it is convenient to express this shift as an action on the spinors. In the four dimensional case, the variable $z$ has an implicit U$(1)$ little group charge. In higher dimensions,  a variable $z$ is chosen which is uncharged under the little group. To compensate, define  matrices parameterizing the choice of shifts as follows:
\begin{align}
M_a{}^b &= - \frac{\lambda_{1a}^{A'} n_{A'A} \lambda_2^{Ab}}{2 p_1 \cdot p_2} , \\
M^a{}_b &= - \frac{\lambda_{2b}^{A'} n_{A'A} \lambda_1^{Aa}}{2 p_1 \cdot p_2} , \\
M_{a'}{}^{b'} &= -\frac{\lambda_{2A}^{b'} n^{AA'} \lambda_{1 A' a'}}{2 p_1 \cdot p_2} , \\
M^{a'}{}_{b'} &= -\frac{ \lambda_{1A}^{a'} n^{A A'} \lambda_{2 A' b'} } {2 p_1 \cdot p_2} \, .
\end{align}
These matrices exist in all even dimensions. However, they are not all independent. In any even dimension, only two of these matrices are independent; which two depends on whether $D=4k$ or $D=4k+2$: see equations \eqref{eq:spinorrelations4k4} and \eqref{eq:spinorrelations4k2}. The definitions are consistent in all cases. An important property of these matrices is that they satisfy
\begin{equation}
M^T M = 0 \, ,
\end{equation}
which ensures that the BCFW shift is linear in $z$. This follows from an easy application of equation \eqref{eq:chiralDiracalg}.

With the matrices $M$ in hand the BCFW shift of the four spinors for particle $1$, namely $\lambda_{1}^{Aa}$, $\lambda_{1a}^{A'}$, $\lambda_1^{Ab}$, and $\lambda_{1b}^{A'}$ reads 
\begin{align}
\label{eq:spinorshifts}
\hat \lambda_{1}^{Aa}(z) &=  \lambda_{1}^{Aa} + z M^{a}{}_{b} \lambda_{2}^{Ab} \\
\hat \lambda_{1a}^{A'}(z) &= \lambda_{1a}^{A'} + z M_{a}{}^{b} \lambda_{2b}^{A'} \\
\hat \lambda_{1 A' a'}(z) &=  \lambda_{1A'a'} + z M_{a'}{}^{b'} \lambda_{2 A'b'} \\
\hat \lambda_{1 A}{}^{a'}(z) &=  \lambda_{1A}{}^{a'} + z M^{a'}{}_{b'} \lambda_{2 A}^{b'} \, .
\end{align}
Again, only half of these shifts are independent. Similarly, the spinors for particle $2$ shift according to
\begin{align}
\label{eq:spinorshiftsII}
\hat \lambda_{2}^{Ab}(z) &=  \lambda_{2}^{Ab} - z \lambda_{1}^{Aa} M^{a}{}_{b}  \\
\hat \lambda_{2b}^{A'}(z) &= \lambda_{2b}^{A'} - z  \lambda_{1a}^{A'} M_{a}{}^{b}\\
\hat \lambda_{2 A' b'}(z) &=  \lambda_{2A'b'} - z \lambda_{1 A'a'} M^{a'}{}_{b'}  \\
\hat \lambda_{2 A}{}^{b'}(z) &=  \lambda_{2A}{}^{b'} - z \lambda_{1 A}^{a'} M_{a'}{}^{b'} \, .
\end{align}
It is now straightforward to check that these shifted spinors correctly reproduce the standard BCFW shift in Eq.~\eqref{eq:standardBCFW}. For example, 
\begin{align}
\hat \lambda_1^{Aa} \hat \lambda^{A'}_{1a} &= p_1^{A A'} + z \left( \lambda_{1a}^{A'} M^a{}_b \lambda_2^{Ab} + \lambda_1^{Aa} M_a{}^b \lambda_{2b}^{A'}\right) + z^2 \lambda_2^{Ab} ( M^T M)_b{}^c \lambda_{2c}^{A'}  \\
&=  p_1^{A A'} - \frac{ z}{s_{12}} \left( p_2^{AB'} n_{B'B} p_1^{BA'} + p_1^{AB'} n_{B'B} p_2^{BA'} \right) \\
&= p_1^{AA'} + z \, n^{AA'} \, .
\end{align}
Note that the nilpotence of the matrices $M$ was necessary to keep the shift linear in $z$.

In a supersymmetric theory, it is very helpful to consider shifts which conserve the supermomentum, i.e., which satisfy
\begin{equation}
\widehat Q_1 + \widehat Q_2 = Q_1 + Q_2 \, .
\end{equation}
Therefore,for the on-shell superspace for which
\begin{equation}
Q_i^{A'}  = \lambda_{ia}^{A'} \eta_{i}^a \, ,
\end{equation}
the supershift involves in addition to \eqref{eq:spinorshifts} and  \eqref{eq:spinorshiftsII} also a shift of the fermionic variables given by
\begin{align}
\hat \eta_{1}^a (z) &= \eta_{1}^a + z M^a{}_b \eta_{2}^b , \\
\hat \eta_{2}^b (z) &= \eta_{1}^a - z \eta_1^a M_a{}^b , \\
\frac{\partial}{\partial \hat \eta_1^a(z)} &= \frac{\partial}{\partial \eta_1^a} + z M_a{}^b  \frac{\partial}{\partial \eta_2^b} \, , \\
\frac{\partial}{\partial \hat \eta_2^b(z)} &= \frac{\partial}{\partial \eta_2^b} - z  \frac{\partial}{\partial \eta_1^a} M^a{}_b \, .
\end{align}
In particular this implies that the conjugate supermomentum is also conserved,
\begin{equation}
\widehat {\bar{Q}}_1 + \widehat {\bar{Q}}_2 = \bar{Q}_1 + \bar{Q}_2 \, .
\end{equation}

\subsection{On-shell recursion}
In general, under a supersymmetric BCFW shift an amplitude becomes a function of a complex parameter $z$. The original amplitude may be obtained by a contour integral around $z=0$, 
\begin{equation}
A(0) = \oint_{z=0} \frac{A(z)}{z} \, .
\end{equation}
As noted by BCFW, pulling the contour to infinity gives
\begin{equation}\label{eq:recurgeneral}
A(0) = - \sum_{z=\textrm{finite}} \textrm{Residues} - \sum_{z=\infty} \textrm{Residue} \, .
\end{equation}
The finite $z$ residues are products of lower point tree amplitudes by tree level unitarity. Hence if the residue at infinity is absent, on-shell recursion relations arise. A sufficient condition for this is if $A(z) \rightarrow 0$ if $z\rightarrow \infty$.

We will make use of  BCFW relations in the context of $\mathcal{N} = 1$ super Yang-Mills theory in 8 dimensions, and type IIB supergravity in 10 dimensions. The dimensional reduction of these theories to four dimensions are well known \cite{Brandhuber:2008pf} \cite{ArkaniHamed:2008gz} to have the property that
\begin{equation}\label{eq:supershiftsthisarticle}
\lim_{z \rightarrow \infty} A^{D=8}(z) \sim \frac{1}{z} \qquad \lim_{z \rightarrow \infty} A^{D=10}(z) \sim \frac{1}{z^2} \, ,
\end{equation}
under a supershift of particles $1$ and $2$. The same property holds in eight and ten dimensions, respectively. Indeed, the $1/z$ behaviour of super Yang-Mills in ten dimensions has already been used in \cite{CaronHuot:2010rj} So let us focus on the gravitational case.

Start by picking the momenta of particles $1$ and $2$, and the shift $n$ to be in some four dimensional subspace. Then our BCFW deformation reduces to the usual four dimensional case. As usual, supersymmetry may be used to pick the polarizations of particles $1$ and $2$ to be whatever we like. Now, all the component amplitudes of the superamplitude have the same $z$ scaling in any dimension as in four dimensions, as can be seen by imagining writing these component amplitudes in terms of Feynman diagrams. It follows that the gravitational amplitude falls off at large $z$ as $1/z^2$. Thus, we may use the usual BCFW recursion in a straightforward manner.

An argument directly in higher dimensions may of course also be set up. For this use the finite supersymmetry transformation generated by
\begin{equation}\label{eq:susytrafo}
\psi_{A} \bar{Q}^A = (c_{a'} 1_{A}^{a'} + d_{a'} 2_A^{a'})  \bar{Q}^A \, ,
\end{equation}
where the coefficients $c$ and $d$ are such that
\begin{equation}\label{eq:detcandd}
\left\{\begin{array}{c}
\eta^a_1 + d_{a'} [2^{a'} 1^a] = 0 \\
\eta^a_2 + c_{a'} [1^{a'} 2^a] = 0 \, ,
\end{array} \right.
\end{equation}
Since as a matrix of $a$ and $a'$
\begin{equation}
\det \left([2^{a'} 1^a] \right) \propto (2 p_1 \cdot p_2 )^{\mathcal{D}/4} \, ,
\end{equation}
holds these equations have a unique solution. The transformation in equation \eqref{eq:susytrafo} is independent of the BCFW shift parameter $z$,
\begin{equation}\label{eq:shiftedsusytrafo}
\psi_{A}= (\hat{c}_{a'} \hat{1}_{A}^{a'} + \hat{d}_{a'} \hat{2}_A^{a'})   =  (c_{a'} 1_{A}^{a'} + d_{a'} 2_A^{a'}) \, .
\end{equation}
To see this, write 
\begin{equation}
\hat{c}_{a'} = c_{a'} + z \tilde{c}_{a'} \qquad \hat{d}_{a'} = d_{a'} + z \tilde{d}_{a'} \, .
\end{equation}
Then from equation \eqref{eq:detcandd}
\begin{equation}
\left\{\begin{array}{c}
M^a{}_b \eta_{2}^b + \tilde{d}_{a'} [2^{a'} 1^a] = 0 \\
- \eta_1^a M_a{}^b + \tilde{c}_{a'} [1^{a'} 2^a] = 0 \, ,
\end{array} \right.
\end{equation}
follows. Plugging the shifted spinor expressions into the left hand side of \eqref{eq:shiftedsusytrafo} and repeated application of  equation \eqref{eq:chiralDiracalg} in conjunction with the just derived equation yield the result. This can then be used to show
\begin{equation}
\lim_{z \rightarrow \infty} A(\{\hat{p}_1, \hat{\eta}_1\}, \{\hat{p}_2, \hat{\eta}_2\}, X ) = \lim_{z \rightarrow \infty} A(\{\hat{p}_1, 0\}, \{\hat{p}_2, 0 \}, \tilde{X} ) \, ,
\end{equation}
where $X$ stands for the quantum numbers of all other particles and $\tilde{X}$ are the same transformed by the finite supersymmetry transformation parametrized by the spinor $\psi_{A}$. Since the transformation is independent of $z$, the only $z$ dependence is now in the first two particles which are the top-states of the supersymmetric multiplet under study.

This general argument can now be applied to yield equation \eqref{eq:supershiftsthisarticle} for the theories of interest here. Note that for IIB sugra one uses that the top-state in any representation scales the same as in any other representation, so one can use the result for the graviton shift to show the above $\frac{1}{z^2}$ behavior.

\subsection*{Solving the Ward identities by on-shell recursion}
A subtlety to do with the finite $z$ residues in equation \eqref{eq:recurgeneral} are the phase conventions for the fermionic integrals arising on cuts. For the shift above the phases on the spinors have to be chosen as
\begin{equation}
\int d\eta_P^{\mathcal{D}/2} A_L(X_L, \{\lambda_P^A, \lambda_P^{A'}, \eta_P \}) A_R(X_R, \{ \lambda_P^A, - \lambda_P^{A'}, \eta_P \}) \, .
\end{equation}
The reason is that the total supermomentum can be split into a `left' and a `right' piece
\begin{align}
Q_{\textrm{tot}} = Q_L + Q_R = \widehat Q_L + \widehat Q_R \\
\bar{Q}_{\textrm{tot}}  = \bar{Q}_L + \bar{Q}_R =  \widehat {\bar{Q}}_L + \widehat {\bar{Q}}_R \, .
\end{align}
so that 
\begin{align}
Q_{\textrm{tot}} A_L A_R & = \left( \widehat Q_L A_L \right) A_R + A_L \left( \widehat Q_R A_R \right) \\
& = \left(\lambda_{a}^{A'} \eta_{P}^a -  \lambda_{a}^{A'} \eta_{P}^a \right) A_L A_R  = 0 \, ,
 \end{align}
and
\begin{align}
\bar{Q}_{\textrm{tot}} \int A_L A_R  & = \int \left( \widehat{\bar{Q}}_L A_L \right) A_R +  A_L \left( \widehat{\bar{Q}}_R A_R \right) \\
& = \int  \lambda_{A,a}   \left[ \frac{d}{d\eta_P^a} \left(A_L\right) A_R +  A_L \frac{d}{d\eta_P^a} \left(A_R\right)\right] \\
& = \int \lambda_{A,a} \frac{d}{d\eta_P^a}   \left[ A_L A_R \right] = 0 \, ,
 \end{align}
hold. The latter is a total derivative which vanishes under the integral. These equations ensure that with the choice of phase as above the terms in the on-shell recursion relations as derived by the super-shift are term-by-term on-shell supersymmetric. Hence the recursion relations solve the supersymmetric Ward identities recursively.

\subsection{Five point amplitudes and beyond}

As an example of the use of on-shell recursion relations, we compute the tree level color ordered five point super Yang-Mills amplitude $\mathcal{A}_5$ in eight dimensions. Let us shift the kinematic data of particles $1$ and $2$. There are two BCFW diagrams; the amplitude is
\begin{equation}
\mathcal{A}_5(1,2,3,4,5) = \frac{i}{s_{15}} \int d^4 \eta_P \mathcal{A}_3(5, \hat 1, -P) \mathcal{A}_4(P, \hat 2, 3, 4) + \frac{1}{s_{23}} \int d^4 \eta_{P'} \mathcal{A}_4(P', 4,5, \hat 1) \mathcal{A}_3(\hat 2, 3, -P') \, .
\end{equation}
Now, in the first term corresponding to the first BCFW diagram, $\mathcal{A}_3(5, \hat 1, -P)$ is annihilated by $Q_5 + \hat Q_1 - \hat Q_P$. Thus one can write $Q_P = \hat Q_1 + Q_5$ in the first term on the right-hand side. Inserting this into the supersymmetric delta function in $\mathcal{A}_4$ a factor $\delta^{(8)}(Q_1 + Q_2 + Q_3 + Q_4 + Q_5)$ arises which may be brought outside the integral. A similar comment holds for the second term on the right hand side. The amplitude becomes
\begin{multline}
\mathcal{A}_5(1,2,3,4,5) = i \delta^{(8)}(Q_1 + Q_2 + Q_3 + Q_4 + Q_5)  \times \\ \left[ \frac{1}{s_{15}s_{\hat 2 3} s_{34}} \int d^4 \eta_P \mathcal{A}_3(5, \hat 1, -P)
 + \frac{1}{s_{\hat 15}s_{2 3} s_{45}} \int d^4 \eta_P \mathcal{A}_3(\hat 2, 3, -P) \right] \, .
\end{multline}
It is very straightforward to compute the integrals inside the square brackets. In fact, we already know these integrals! They just correspond to extracting all the component amplitudes of the three point superamplitude, in which the intermediate propagating particle is treated as a scalar. Thus, for example,
\begin{equation}
\int d^4 \eta_P \mathcal{A}_3(5, \hat 1, -P) = \left(\eta_5^a  u(5)_a^{j'} + \hat \eta_1^a u(1)_a^{j'} \right) \epsilon_{j'k'} \left( u(5)_b^{k'}  \eta_5^b + \hat u(1)^{k'}_b \eta_1^b \right) \, .
\end{equation} 
It follows that the five point amplitude is simply
\begin{multline}
\mathcal{A}_5(1,2,3,4,5) = i \delta^{(8)}(Q_1 + Q_2 + Q_3 + Q_4 + Q_5)  \times  \\ \left[ \frac{  \left(\eta_5 \cdot u_5 + \hat \eta_1 \cdot u_1 \right)^{j'} \epsilon_{j' k'}  \left( u_5 \cdot \eta_5 + u_1 \cdot \hat \eta_1 \right)^{k'} }{s_{15}s_{\hat 2 3} s_{34}} 
 + \frac{ \left(\hat \eta_2 \cdot u_2 + \eta_3 \cdot u_3 \right)^{j'} \epsilon_{j' k'}  \left( u_2 \cdot \hat \eta_2 + u_3 \cdot \eta_3 \right)^{k'}}{s_{\hat 15}s_{2 3} s_{45}}  \right]  \, ,
\end{multline}
where the inner products are the natural inner products in the spinor spaces. Notice that this equation is an eight dimensional generalization of the six dimensional equation 3.7 discussed in \cite{Bern:2010qa}. 

It is possible in principle to move beyond the just presented five point example and calculate higher order amplitudes, both in super Yang-Mills theory in eight dimensions as well as IIB sugra in ten. The six point gauge theory calculation for instance would contain three terms, all of which are known in fairly compact form.


\section{Discussion and conclusions}
The engine of this paper is a formulation of spinor-helicity which works in general dimensions, and which unifies and extends previous work in this direction, e.g. \cite{Cheung:2009dc}, \cite{Boels:2009bv} and  \cite{CaronHuot:2010rj}. These methods were applied to the problem of constructing on-shell superspaces in these dimensions. The role the various groups (Poincare, little, tiny, super) play has been highlighted throughout. We have shown that the Ward identities of the supersymmetry algebra may be solved recursively in on-shell superspaces constructed using supersymmetric higher dimensional on-shell recursion. Explicit example applications of the technology have been presented in eight and ten dimensions for maximal super Yang-Mills as well as IIB sugra respectively. 

There are various directions for interesting future research which lead from this article. One is the relation of our techniques and results to off-shell methods in string and field theory,  especially those with manifest (target space) supersymmetry. For instance, the scattering amplitude in IIB superstring theory has been obtained in \cite{Policastro:2006vt} using pure spinor methods. It would be interesting to see if there is a natural interplay between these and the on-shell spinors considered here especially as on-shell spinors are pure \cite{Boels:2009bv}. Perhaps the technically easiest approach here is through the light-cone superspace, see e.g. \cite{Metsaev:2004wv}. 

A separate direction is the calculation of loop amplitudes in highly supersymmetric theories. As a proof of concept application of the technology developed here a particular `sewing relation' useful for iterated unitarity cuts in $\mathcal{N}=8$ supergravity calculations is proven in appendix \ref{app:proofsewing}.

Note that we imposed no restriction on the number of dimensions in the general formalism developed in the body of the paper: in principle superfields in any dimension may be constructed, along with supersymmetric delta functions. However, standard lore is that, with some modest assumptions, interacting unitary quantum field theories may not be constructed in dimensions higher than eleven. It would be interesting to see this constraint arising within purely on-shell methods. It would also be of interest to explore the space of theories with non-trivial three point functions using our general techniques, along the lines of the work of~\cite{Huang:2010qy,Czech:2011dk}.

\section*{Acknowledgments}
RB would like to thank the Asia Pacific Center for Theoretical Physics for the workshop ``Strings, Branes and Liouville Theory 6'' which inspired the present article. This work was supported by the German Science Foundation (DFG) within the Collaborative Research Center 676 ``Particles, Strings and the Early Universe''. 

\vspace{1cm}

\appendix


\section{On-shell proof of the four particle `sewing relation' in type IIB sugra}\label{app:proofsewing}
This appendix contains a short proof of the so-called sewing relation, equation (5.6) in \cite{Bern:1998ug}, using the ten dimensional super-amplitude. This formula arises in the study of $2$ particle cuts of sugra loop amplitudes. The reader is referred to that paper for a discussion of its uses. The relation reads for the graviton tree amplitudes
\begin{multline}\label{eq:sewingrelation}
\sum_{\mathcal{N} =8 \textrm{ states}} A_4(-l_1, 1, 2, l_2) A_4(-l_2, 3, 4, l_1)  = \\
 (s\, t\, u) A_4(1,2,3,4) \left[\frac{1}{(l_1-k_1)^2} + \frac{1}{ (l_1-k_2)^2} \right]  \left[\frac{1}{(l_2-k_3)^2} + \frac{1}{ (l_2-k_4)^2} \right] \, ,
\end{multline}
where the sum ranges over the content of the $\mathcal{N}=8$ maximal supergravity in four dimensions in the two cut legs. All momenta in this formula are on-shell. It was originally proven in all dimensions through use of the KLT relations combined with a component form analysis of the Yang-Mills amplitudes. Here it is shown superfields give a compact derivation. 

In terms of the superamplitude the sum over states on the right hand side of the above relation can be replaced by two fermionic integrations,
\begin{multline}
\sum_{\mathcal{N} =8 \textrm{ states}} A_4(-l_1, 1, 2, l_2) A_4(-l_2, 3, 4, l_1)  \rightarrow \\ \int (d\eta_1)^8 (d\eta_2)^8 A_4(\{- l_1, \eta_1\},  1, 2, \{l_2, \eta_2\}) A_4(\{-l_2,\eta_2\}, 3, 4, \{ l_1,\eta_1\}) \, ,
\end{multline} 
where the fermionic variables of the legs $1,2,3,4$ have been suppressed. Inserting the superamplitudes and extracting the part of the amplitude independent of the fermionic coordinates yields
\begin{multline}
 \int  (d\eta_1)^8 (d\eta_2)^8 A_4(\{- l_1, \eta_1\},  1, 2, \{l_2, \eta_2\}) A_4(\{-l_2,\eta_2\}, 3, 4, \{ l_1,\eta_1\})  = \\ \frac{1}{s^4}  \left[\frac{1}{(l_1-k_1)^2} + \frac{1}{ (l_1-k_2)^2} \right]  \left[\frac{1}{(l_2-k_3)^2} + \frac{1}{ (l_2-k_4)^2} \right] \\ \int  (d\eta_1)^8 (d\eta_2)^8 \delta^{16}\left(Q_{-l_1} + Q_{1} + Q_2 + Q_{l_2} \right) \delta^{16}\left( Q_{-l_2} + Q_{3} + Q_4 + Q_{l_1} \right) \, .
 \end{multline}
Noticing that
\begin{multline}
 \int  (d\eta_1)^8 (d\eta_2)^8 \delta^{16}\left(Q_{-l_1} + Q_{1} + Q_2 + Q_{l_2} \right) \delta^{16}\left( Q_{-l_2} + Q_{3} + Q_4 + Q_{l_1} \right)   = \\
  =  \delta^{16}\left( Q_{1} + Q_{2} + Q_3 + Q_{4} \right) \,\, \int  (d\eta_1)^8 (d\eta_2)^8\delta^{16}\left(- Q_{l_1} + Q_{l_2} \right) \, ,
\end{multline}
and plugging this into the previous equation yields 
\begin{multline}
 \int  (d\eta_1)^8 (d\eta_2)^8 A_4(\{- l_1, \eta_1\},  1, 2, \{l_2, \eta_2\}) A_4(\{-l_2,\eta_2\}, 3, 4, \{ l_1,\eta_1\})  = \\ \frac{1}{s^4}  \left[\frac{1}{(l_1-k_1)^2} + \frac{1}{ (l_1-k_2)^2} \right]  \left[\frac{1}{(l_2-k_3)^2} + \frac{1}{ (l_2-k_4)^2} \right] \\ (s t u) A_4(1,2,3,4)   \int  (d\eta_1)^8 (d\eta_2)^8\delta^{16}\left(- Q_{l_1} + Q_{l_2} \right) \, ,
 \end{multline}
Since
\begin{equation}
\frac{1}{s^4} \int  (d\eta_1)^8 (d\eta_2)^8  \delta^{16}\left(- Q_{l_1} + Q_{l_2} \right) = 1 \, ,
\end{equation}
follows by equation \eqref{eq:spinorsleadingtoinnprod} equation \eqref{eq:sewingrelation} holds. Since the driving relation in this derivation is simply the integral over the fermionic delta functions, the same calculation may be repeated at any loop order as is needed in \cite{Bern:1998ug} for instance. 


\section{Three in six}\label{app:threeinsix}
In this appendix, a form of the three point superamplitude in six dimensions will calculated. An equivalent expression (in addition to higher point amplitudes) was described previously by~\cite{Dennen:2009vk}. 

In six dimensions for $\mathcal{N}=(1,1)$ supersymmetry there are two supercharges to be considered which will be represented as
\begin{align}
Q^A = \eta_a \lambda^{A,a} \\
Q_{A'} = \eta^{a'} \lambda_{A',a'} \, ,
\end{align}
and their natural conjugates. Since up and down chiral spinor indices are independent in six dimensions this notation for the two supercharges is unambiguous. By the general argument in the main text the three point amplitude is proportional to
\begin{equation}
A_{susy} = C  \left( \epsilon_{ABCD} Q^{A} Q^{B} Q^{C} \xi^{D}_{a'}\right) \, \left( \epsilon^{A'B'C'D'} Q_{A'} Q_{B'} Q_{C'} \xi_{D'}^{a}\right) L_{a}{}^{a'} \, ,
\end{equation}
for some proportionality constant C. To show this and to calculate the proportionality constant, let us calculate the scalar-gluon-anti-scalar amplitude. In terms of $u$ variables this is by equation \eqref{eq:uversionthreepointamp}
\begin{equation}
A(\phi g^{b}_{b'} \bar{\phi}) = 2 u_{2}^{b}\, u_{2,b'}  \, .
\end{equation}
This component amplitude follows from integrating over the superamplitude as
\begin{equation}
A(\phi g^b_{b'} \bar{\phi}) = \int d\eta_{2,b} d\eta_2^{b'} d\eta_3^4 A_{susy} \, .
\end{equation}
This integral is easy to perform and leads to
\begin{align}
A(\phi g^b_{b'} \bar{\phi}) & = C \frac{1}{(2!)^2} \left( \epsilon_{ABCD} 3_{a}^{A} 3^{B,a} 2^{C,b} \xi^{D}_{a'}\right) \, \left( \epsilon^{A'B'C'D'} 3_{a',A'} 3_{B'}^{a'} 2_{C',b'} \xi_{D'}^{a}\right) L_{a}{}^{a'} \\
& = C  \left(  3_{a',A} 3^{a'}_B 2^{A,b} \xi^{B}_{a'}\right) \, \left(3_{a}^{A'} 3^{B',a} 2_{A',b'} \xi_{B;}^{a}\right) L_{a}{}^{a'} \\
& = - C  \left(  1_{a',A} 1^{a'}_B 2^{A,b} \xi^{B}_{a'}\right) \, \left(3_{a'}^{A'} 3^{B',a'} 2_{C',b'} \xi_{D}^{a}\right) L_{a}{}^{a'} \, .
\end{align}
Converting to u's then gives
\begin{equation}
A(\phi g^b_{b'} \bar{\phi}) = - C \,  u_2^b \, u_{2,b'} \left( m_1 m_3 u_{1,a'} u_3^{a} L_{a}{}^{a'} \right) \, .
\end{equation}
Hence for
\begin{equation}
C  = - \frac{2}{ L^{a}{}_{a'}  [\xi_a | \slash \!\!\! 1 \slash \!\!\! 3 |\xi^{a'} ] } \, .
\end{equation}
The superamplitude reproduces the right component amplitude, taking into account the special properties of six dimensional spinors.

In conclusion, the supersymmetric three point function in $\mathcal{N}=(1,1)$ supersymmetric Yang-Mills theory in six dimensions is given by
\begin{equation}
A_{susy} = \frac{2}{ L^{a}{}_{a'}  [\xi_a | \slash \!\!\! 1 \slash \!\!\! 3 |\xi^{a'} ] } \left( \epsilon_{ABCD} Q^{A} Q^{B} Q^{C} \xi^{D}_{a'}\right) \, \left( \epsilon^{A'B'C'D'} Q_{A'} Q_{B'} Q_{C'} \xi_{D}^{a}\right) L_{a}{}^{a'} \, ,
\end{equation}
as long as
\begin{equation}
L^{a}{}_{a'}  \langle1_{a}3^{b'} \rangle \neq 0 \, .
\end{equation}

\bibliographystyle{JHEP}

\bibliography{10Dbib}
\end{document}